\documentclass{article}%
\usepackage{amsmath}
\usepackage{amsfonts}
\usepackage{amssymb}
\usepackage{graphicx}
\usepackage{fullpage}%
\newtheorem{theorem}{Theorem}

\newtheorem{corollary}[theorem]{Corollary}

\newtheorem{definition}[theorem]{Definition}

\newtheorem{lemma}[theorem]{Lemma}

\newtheorem{proposition}[theorem]{Proposition}

\newenvironment{proof}[1][Proof]{\noindent\textbf{#1.} }{\ \rule{0.5em}{0.5em}}
\begin{document}

\title{Oracles Are Subtle But Not Malicious}
\author{Scott Aaronson\thanks{Email: aaronson@ias.edu. \ This research was done while
the author was a postdoc at the Institute for Advanced Study in Princeton,
supported by an NSF grant.}}
\date{}
\maketitle

\begin{abstract}
Theoretical computer scientists have been debating the role of oracles since
the 1970's. \ This paper illustrates both that oracles can give us nontrivial
insights about the barrier problems in circuit complexity, and that they need
not prevent us from trying to solve those problems.

First, we give an oracle relative to which $\mathsf{PP}$ has linear-sized
circuits, by proving a new lower bound for perceptrons and low-degree
threshold polynomials. \ This oracle settles a longstanding open question, and
generalizes earlier results due to Beigel and to Buhrman, Fortnow, and
Thierauf. \ More importantly, it implies \textit{the first nonrelativizing
separation of \textquotedblleft traditional\textquotedblright\ complexity
classes}, as opposed to interactive proof classes such as $\mathsf{MIP}$\ and
$\mathsf{MA}_{\mathsf{EXP}}$. \ For Vinodchandran showed, by a nonrelativizing
argument, that $\mathsf{PP}$\ does not have circuits of size $n^{k}$\ for any
fixed $k$. \ We present an alternative proof of this fact, which shows that
$\mathsf{PP}$\ does not even have quantum circuits of size $n^{k}$ with
quantum advice. \ To our knowledge, this is the first nontrivial lower bound
on quantum circuit size.

Second, we study a beautiful algorithm of Bshouty et al. for learning Boolean
circuits in $\mathsf{ZPP}^{\mathsf{NP}}$. \ We show that the $\mathsf{NP}%
$\ queries in this algorithm cannot be parallelized by any relativizing
technique, by giving an oracle relative to which $\mathsf{ZPP}_{||}%
^{\mathsf{NP}}$\ and even $\mathsf{BPP}_{||}^{\mathsf{NP}}$ have linear-size
circuits. \ On the other hand, we also show that the $\mathsf{NP}$\ queries
\textit{could} be parallelized if $\mathsf{P}=\mathsf{NP}$. \ Thus, classes
such as $\mathsf{ZPP}_{||}^{\mathsf{NP}}$\ inhabit a \textquotedblleft
twilight zone,\textquotedblright\ where we need to distinguish between
relativizing and black-box techniques. \ Our results on this subject have
implications for computational learning theory as well as for the circuit
minimization problem.

\end{abstract}

\section{Introduction\label{INTRO}}

It is often lamented that, half a century after Shannon's insight
\cite{shannon} that almost all Boolean functions require exponential-size
circuits, there is still no explicit function for which we can prove even a
superlinear lower bound. \ Yet whether this lament is justified depends on
what we mean by \textquotedblleft explicit.\textquotedblright\ \ For in 1982,
Kannan \cite{kannan} did show that for every constant $k$, there exists a
language in $\mathsf{\Sigma}_{2}^{p}$\ (the second level of the polynomial
hierarchy)\ that does not have circuits of size $n^{k}$. \ His proof used the
oldest trick in the book: \textit{diagonalization}, defined broadly as any
method for simulating all machines in one class by a single machine in
another. \ In some sense, diagonalization is still the only method we know
that zeroes in on a specific property of the function being lower-bounded, and
thereby escapes the jaws of Razborov and Rudich \cite{rr}.

But can we generalize Kannan's theorem to other complexity classes? \ A decade
ago, Bshouty et al. \cite{bshouty}\ discovered an algorithm to learn Boolean
circuits in $\mathsf{ZPP}^{\mathsf{NP}}$ (that is, probabilistic polynomial
time with $\mathsf{NP}$\ oracle). \ As noticed by K\"{o}bler and Watanabe
\cite{kobler}, the existence of this algorithm implies that $\mathsf{ZPP}%
^{\mathsf{NP}}$ itself cannot have circuits of size $n^{k}$\ for any
$k$.\footnote{For Bshouty et al.'s algorithm implies the following improvement
to the celebrated Karp-Lipton theorem \cite{kl}: if $\mathsf{NP}%
\subset\mathsf{P/poly}$\ then $\mathsf{PH}$\ collapses to $\mathsf{ZPP}%
^{\mathsf{NP}}$. \ There are then two cases: if $\mathsf{NP}\not \subset
\mathsf{P/poly}$, then certainly $\mathsf{ZPP}^{\mathsf{NP}}\not \subset
\mathsf{P/poly}$\ as well and we are done. \ On the other hand, if
$\mathsf{NP}\subset\mathsf{P/poly}$, then $\mathsf{ZPP}^{\mathsf{NP}%
}=\mathsf{PH}$, but we already know from Kannan's theorem that $\mathsf{PH}%
$\ does not have circuits of size $n^{k}$. \ Indeed, we can repeat this
argument for the class $\mathsf{S}_{2}^{p}$, which Cai \cite{cai}\ showed is
contained in $\mathsf{ZPP}^{\mathsf{NP}}$.}

So our task as lowerboundsmen and lowerboundswomen seems straightforward:
namely, to find increasingly powerful algorithms for learning Boolean
circuits, which can then be turned around to yield increasingly powerful
circuit lower bounds. \ But when we try to do this, we quickly run into the
brick wall of \textit{relativization}. \ Just as Baker, Gill, and Solovay
\cite{bgs}\ gave a relativized world where $\mathsf{P}=\mathsf{NP}$, so Wilson
\cite{wilson}\ gave relativized worlds where $\mathsf{NP}$\ and $\mathsf{P}%
^{\mathsf{NP}}$\ have linear-size circuits. \ Since the results of Kannan
\cite{kannan} and Bshouty et al. \cite{bshouty}\ relativize, this suggests
that new techniques will be needed to make further progress.

Yet attitudes toward relativization vary greatly within our community. \ Some
computer scientists ridicule oracle results as elaborate formalizations of the
obvious---apparently believing that (1) there exist relativized worlds where
just about anything is true, (2) the creation of such worlds is a routine
exercise, (3) the only conjectures ruled out by oracle results are trivially
false ones, which no serious researcher would waste time trying to prove, and
(4) nonrelativizing results such as $\mathsf{IP}=\mathsf{PSPACE}%
$\ \cite{shamir} render oracles irrelevant anyway. \ At the other extreme,
some computer scientists see oracle results not as a spur to create
nonrelativizing techniques or as a guide to where such techniques might be
needed, but as an excuse to abandon hope.

This paper will offer new counterexamples to both of these views, in the
context of circuit lower bounds. \ We focus on two related topics: first, the
classical and quantum circuit complexity of $\mathsf{PP}$; and second, the
learnability of Boolean circuits using parallel $\mathsf{NP}$\ queries.

\subsection{On \textsf{PP} and Quantum Circuits\label{PPINTRO}}

In Section \ref{PPORACLE},\ we give an oracle relative to which $\mathsf{PP}%
$\ has linear-size circuits. \ Here $\mathsf{PP}$\ is the class of languages
accepted by a nondeterministic polynomial-time Turing machine that accepts and
if and only if the majority of its paths do. \ Our construction also yields an
oracle relative to which $\mathsf{PEXP}$\ (the exponential-time version of
$\mathsf{PP}$) has polynomial-size circuits, and indeed $\mathsf{P}%
^{\mathsf{NP}}=\mathsf{\oplus P}=\mathsf{PEXP}$. \ This settles several
questions that were open for years,\footnote{Lance Fortnow, personal
communication.} and subsumes at least three previous results:

\begin{enumerate}
\item[(1)] that of Beigel \cite{beigel}\ giving an oracle relative to which
$\mathsf{P}^{\mathsf{NP}}\not \subset \mathsf{PP}$ (since clearly
$\mathsf{P}^{\mathsf{NP}}=\mathsf{PEXP}$\ implies $\mathsf{P}^{\mathsf{NP}%
}\not \subset \mathsf{PP}$),

\item[(2)] that of Buhrman, Fortnow, and Thierauf \cite{bft}\ giving an oracle
relative to which $\mathsf{MA}_{\mathsf{EXP}}\subset\mathsf{P/poly}$, and

\item[(3)] that of Buhrman et al. \cite{bfft}\ giving an oracle relative to
which $\mathsf{P}^{\mathsf{NP}}=\mathsf{NEXP}$.
\end{enumerate}

Note that our result is nearly optimal, since Toda's theorem \cite{toda}
yields a relativizing proof\ that $\mathsf{P}^{\mathsf{PP}}$\ and even
$\mathsf{BP\cdot PP}$ do not have circuits of any fixed polynomial size.

Our proof first represents each $\mathsf{PP}$\ machine by a low-degree
multilinear polynomial, whose variables are the bits of the oracle string.
\ It then combines these polynomials into a single polynomial called $Q$.
\ The key fact is that, if there are no variables left \textquotedblleft
unmonitored\textquotedblright\ by the component polynomials, then we can
modify the oracle in a way that increases $Q$. \ Since $Q$ can only increase a
finite number of times, it follows that we will eventually win our
\textquotedblleft war of attrition\textquotedblright\ against the polynomials,
at which point we can simply write down what each machine does in an
unmonitored part of the oracle string. \ The main novelty of the proof lies in
how we combine the polynomials to create $Q$.

We can state our result alternatively in terms of \textit{perceptrons}
\cite{mp}, also known as threshold-of-AND circuits or polynomial threshold
functions. \ Call a perceptron \textquotedblleft small\textquotedblright\ if
it has size $2^{n^{o\left(  1\right)  }}$, order $n^{o\left(  1\right)  }$,
and weights in $\left\{  -1,1\right\}  $.\ \ Also, given an $n$-bit string
$x_{1}\ldots x_{n}$, recall that the ODDMAXBIT problem is to decide whether
the maximum $i$ such that $x_{i}=1$\ is even or odd, promised that such an $i$
exists.\ Then Beigel \cite{beigel}\ showed that no small perceptron can solve
ODDMAXBIT. \ What we show is a strong generalization of Beigel's theorem: for
any $k=n^{o\left(  1\right)  }$\ small perceptrons, there exists a
\textquotedblleft problem set\textquotedblright\ consisting of $k$ ODDMAXBIT
instances, such that for every $i$,\ the $i^{th}$\ perceptron will get the
$i^{th}$\ problem wrong even if it can examine the whole problem set.
\ Previously this had been open even for $k=2$.

But the real motivation for our result is that in the unrelativized world,
$\mathsf{PP}$\ is known \textit{not} to have linear-size circuits. \ Indeed,
Vinodchandran \cite{vinodchandran} showed that for every $k$, there exists a
language in $\mathsf{PP}$\ that does not have circuits of size $n^{k}$. \ As a
consequence, we obtain \textit{the first nonrelativizing separation that does
not involve artificial classes or classes defined using interactive proofs}.
\ There have been nonrelativizing separations in the past, but most of them
have followed easily from the collapse of interactive proof classes: for
example, $\mathsf{NP}\neq\mathsf{MIP}$ from $\mathsf{MIP}=\mathsf{NEXP}$
\cite{bfl}, and $\mathsf{IP}\not \subset \mathsf{SIZE}\left(  n^{k}\right)
$\ from $\mathsf{IP}=\mathsf{PSPACE}$\ \cite{shamir}. \ The one exception was
the result of Buhrman, Fortnow, and Thierauf \cite{bft}\ that $\mathsf{MA}%
_{\mathsf{EXP}}\not \subset \mathsf{P/poly}$, where $\mathsf{MA}%
_{\mathsf{EXP}}$\ is the exponential-time version of $\mathsf{MA}$. \ However,
the class $\mathsf{MA}_{\mathsf{EXP}}$\ exists for the specific purpose of not
being contained in $\mathsf{P/poly}$,\ and the resulting separation does not
scale down below $\mathsf{NEXP}$, to show (for example) that $\mathsf{MA}$
does not have linear-size circuits. \ By contrast, $\mathsf{PP}$\ is one of
the most natural complexity classes there is. \ That is why, in our opinion,
our result adds some heft to the idea that \textit{currently-understood
nonrelativizing techniques can lead to progress on the fundamental questions
of complexity theory.}

The actual lower bound of Vinodchandran \cite{vinodchandran}\ follows easily
from three well-known results: the LFKN interactive protocol for the permanent
\cite{lfkn}, Toda's theorem \cite{toda}, and Kannan's theorem \cite{kannan}%
.\footnote{Suppose by contradiction that $\mathsf{PP}$\ has circuits of size
$n^{k}$. \ Then $\mathsf{P}^{\mathsf{\#P}}\subset\mathsf{P/poly}$, and
therefore $\mathsf{MA}=\mathsf{PP}=\mathsf{P}^{\mathsf{\#P}}$\ by a result of
LFKN \cite{lfkn}\ (this is the only part of the proof that fails to
relativize). \ Now $\mathsf{MA}\subseteq\mathsf{\Sigma}_{2}^{p}\subseteq
\mathsf{P}^{\mathsf{\#P}}$\ by Toda's theorem \cite{toda}, so $\mathsf{\Sigma
}_{2}^{p}=\mathsf{PP}$\ as well. \ But we already know from Kannan's theorem
\cite{kannan}\ that $\mathsf{\Sigma}_{2}^{p}$\ does not have circuits of size
$n^{k}$.} \ In Section \ref{QUANTUM}, we present an alternative, more
self-contained proof, which does not go through Toda's theorem. \ As a bonus,
our proof also shows that $\mathsf{PP}$\ does not have \textit{quantum}
circuits of size $n^{k}$\ for any $k$. \ Indeed, this remains true even if the
quantum circuits are given \textquotedblleft quantum advice
states\textquotedblright\ on $n^{k}$\ qubits,\ which might require exponential
time to prepare. \ One part of our proof is a \textquotedblleft quantum
Karp-Lipton theorem,\textquotedblright\ which states that if $\mathsf{PP}$ has
polynomial-size quantum circuits, then the \textquotedblleft counting
hierarchy\textquotedblright\ (consisting of $\mathsf{PP}$, $\mathsf{PP}%
^{\mathsf{PP}}$, $\mathsf{PP}^{\mathsf{PP}^{\mathsf{PP}}}$, and so on)
collapses to $\mathsf{QMA}$, the quantum analogue of $\mathsf{NP}$. \ By
analogy to the classical nonrelativizing separation of Buhrman, Fortnow, and
Thierauf \cite{bft}, we also show that $\mathsf{QMA}_{\mathsf{EXP}}$, the
exponential-time version of $\mathsf{QMA}$, is not contained in
$\mathsf{BQP/qpoly}$. \ Indeed, $\mathsf{QMA}_{\mathsf{EXP}}$\ requires
quantum circuits of at least \textquotedblleft
half-exponential\textquotedblright\ size, meaning size $f\left(  n\right)
$\ where $f\left(  f\left(  n\right)  \right)  $\ grows
exponentially.\footnote{See Miltersen, Vinodchandran, and Watanabe
\cite{mvw}\ for a discussion of this concept.}

While none of the results in Section \ref{QUANTUM}\ are really difficult,\ we
include them here for three reasons:

\begin{enumerate}
\item[(1)] So far as we know, the only existing lower bounds for arbitrary
quantum circuits are due to Nishimura and Yamakami \cite{ny}, who showed
(among other things) that $\mathsf{EESPACE}\not \subset \mathsf{BQP/qpoly}%
$.\footnote{A similar bound is implicit in a paper by Stockmeyer and Meyer
\cite{sm}.} \ We felt it worthwhile to point out that much better bounds are possible.

\item[(2)] When it comes to understanding the limitations of quantum
computers, our knowledge to date consists almost entirely of oracle lower
bounds. \ Many researchers have told us that they would much prefer to see
some unrelativized results, or at the very least conditional statements---for
example, \textquotedblleft if $\mathsf{NP}$-complete problems\ are solvable in
quantum polynomial time, then the polynomial hierarchy
collapses.\textquotedblright\ \ The results of Section \ref{QUANTUM} represent
a first step in that direction.

\item[(3)] Recently Aaronson \cite{aar:pp}\ gave a new characterization of
$\mathsf{PP}$, as the class of problems solvable in quantum polynomial time,
given the ability to \textit{postselect} (that is, to discard all runs of the
computation in which a given measurement result does not occur). \ If we
replace \textquotedblleft quantum\textquotedblright\ by \textquotedblleft
randomized\textquotedblright\ in this definition,\ then we obtain a classical
complexity class called $\mathsf{BPP}_{\mathsf{path}}$, which was introduced
by Han, Hemaspaandra, and Thierauf \cite{hht}. \ So the fact that we can prove
a quantum circuit lower bound for $\mathsf{PP}$\ implies one of two things:
either that (i) we can prove a nonrelativizing quantum separation theorem, but
\textit{not} the classical analogue of the same theorem, or that (ii) we
should be able to prove classical circuit lower bound for $\mathsf{BPP}%
_{\mathsf{path}}$. \ As we will see later, the latter possibility would be a
significant breakthrough.
\end{enumerate}

\subsection{On Parallel $\mathsf{NP}$ Queries and Black-Box
Learning\label{PARINTRO}}

In a second part of the paper, we study the learning algorithm of Bshouty et
al. \cite{bshouty}\ mentioned earlier. \ Given a Boolean function $f$\ that is
promised to have a polynomial-size circuit, this algorithm \textit{finds} such
a circuit in the class $\mathsf{ZPP}^{\mathsf{NP}^{f}}$: that is, zero-error
probabilistic polynomial time with $\mathsf{NP}$ oracle with oracle for $f$.
\ One of the most basic questions about this algorithm is whether the
$\mathsf{NP}$ queries can be made nonadaptive. \ For if so, then we
immediately obtain a new circuit lower bound: namely that $\mathsf{ZPP}%
_{||}^{\mathsf{NP}}$\ (that is, $\mathsf{ZPP}$ with parallel\ $\mathsf{NP}%
$\ queries) does not have circuits of size $n^{k}$\ for any $k$.\footnote{This
follows from the same reasoning used by K\"{o}bler and Watanabe \cite{kobler}
to show that $\mathsf{ZPP}^{\mathsf{NP}}$\ does not have circuits of size
$n^{k}$. \ For such an algorithm would readily imply that if $\mathsf{NP}%
\subset\mathsf{P/poly}$, then $\mathsf{PH}$\ collapses to $\mathsf{ZPP}%
_{||}^{\mathsf{NP}}$.} \ Conceptually, this would not be so far from showing
that $\mathsf{NP}$\ itself does not have circuits of size $n^{k}%
$.\footnote{For as observed by Shaltiel and Umans \cite{shaltielumans}\ and
Fortnow and Klivans \cite{fk}\ among others, there is an intimate connection
between the classes $\mathsf{P}_{||}^{\mathsf{NP}}$\ and $\mathsf{NP/l{}og}$.
\ Furthermore, any circuit lower bound for $\mathsf{NP/l{}og}$ implies the
same lower bound for $\mathsf{NP}$, since we can tack the advice onto the
input.}

Let $\mathcal{C}$ be the set of circuits of size $n^{k}$. \ In Bshouty et
al.'s algorithm, we repeatedly ask the $\mathsf{NP}$\ oracle to find us an
input $x_{t}$ such that, \textit{among the circuits in }$\mathcal{C}$\textit{
that succeed on all previous inputs }$x_{1},\ldots,x_{t-1}$, at least a
$1/3$\ fraction fail on $x_{t}$. \ Since each such input reduces the number of
circuits \textquotedblleft still in the running\textquotedblright\ by at least
a constant factor, this process can continue for at most $\log\left\vert
\mathcal{C}\right\vert $\ steps. \ Furthermore, when it ends, by assumption we
have a set $\mathcal{C}^{\ast}$\ of circuits such that for all inputs $x$, a
uniform random circuit drawn from $\mathcal{C}^{\ast}$\ will succeed on $x$
with probability at least $2/3$. \ So now all we have to do is sample a
polynomial number of circuits from $\mathcal{C}^{\ast}$, then generate a new
circuit that outputs the majority answer among the sampled circuits. \ The
technical part is to express the concepts \textquotedblleft at least a $1/3$
fraction\textquotedblright\ and \textquotedblleft a uniform random
sample\textquotedblright\ in $\mathsf{NP}$. \ For that Bshouty et al. use
pairwise-independent hash functions.

When we examine the above algorithm, it is far from obvious that adaptive
$\mathsf{NP}$\ queries are necessary. \ For why can't we simply ask the
following question in parallel, for all $T\leq\log\left\vert \mathcal{C}%
\right\vert $?

\begin{quote}
\textquotedblleft Do there exist inputs $x_{1},\ldots,x_{T}$, such that at
least a $1/3$\ fraction of circuits in $\mathcal{C}$\ fail on $x_{1}$, and
among the circuits that succeed on $x_{1}$, at least a $1/3$\ fraction fail on
$x_{2}$, and among the circuits that succeed on $x_{1}$ and $x_{2}$, at least
a $1/3$\ fraction fail on $x_{3}$, \ldots\ and so on up to $x_{T}%
$?\textquotedblright
\end{quote}

By making clever use of hashing and approximate counting, perhaps we could
control the number of circuits that succeed on $x_{1},\ldots,x_{t}$\ for all
$t\leq T$. \ In that case, by finding the largest $T$ such that the above
question returns a positive answer, and then applying the Valiant-Vazirani
reduction \cite{vv}\ and other standard techniques, we would achieve the
desired parallelization of Bshouty et al.'s algorithm. \ Indeed, when we began
studying the topic, it seemed entirely likely to us that this was possible.

Nevertheless, in Section \ref{PARORACLE}\ we give an oracle relative to which
$\mathsf{ZPP}_{||}^{\mathsf{NP}}$\ and even $\mathsf{BPP}_{||}^{\mathsf{NP}}%
$\ have linear-size circuits. \ The overall strategy of our oracle
construction is the same as for $\mathsf{PP}$, but the details are somewhat
less elegant. \ The existence of this oracle means that any parallelization of
Bshouty et al.'s algorithm will need to use nonrelativizing techniques.

Yet even here, the truth is subtler than one might imagine. \ To explain why,
we need to distinguish carefully between \textit{relativizing} and
\textit{black-box} algorithms. \ An algorithm for learning Boolean circuits is
\textit{relativizing}\ if, when given access to an oracle $A$, the algorithm
can learn circuits that are also given access to $A$. \ But a nonrelativizing
algorithm can still be \textit{black-box}, in the sense that it learns about
the target function $f$ only by querying it, and does not exploit any succinct
description of $f$ (for example, that $f\left(  x\right)  =1$\ if and only if
$x$ encodes a satisfiable Boolean formula). \ Bshouty et al.'s algorithm is
both relativizing \textit{and} black-box. \ What our oracle construction shows
is that\ no relativizing algorithm can learn Boolean circuits in
$\mathsf{BPP}_{||}^{\mathsf{NP}}$. \ But what about a nonrelativizing yet
still black-box algorithm?

Surprisingly, we show in Section \ref{ALG}\ that if $\mathsf{P}=\mathsf{NP}$,
then there \textit{is} a black-box algorithm to learn Boolean circuits even in
$\mathsf{P}_{||}^{\mathsf{NP}}$ (as well as in $\mathsf{NP/l{}og}$). \ Despite
the outlandishness of the premise, this theorem is not trivial, and requires
many of the same techniques originally used by Bshouty et al. \cite{bshouty}.
\ One way to interpret the theorem is that we cannot show the
\textit{impossibility} of black-box learning in $\mathsf{P}_{||}^{\mathsf{NP}%
}$, without also showing that $\mathsf{P}\neq\mathsf{NP}$. \ By contrast, it
is easy to show that black-box learning is impossible in $\mathsf{NP}$,
regardless of what computational assumptions we make.\footnote{Note that by
\textquotedblleft learn,\textquotedblright\ we always mean \textquotedblleft
learn exactly\textquotedblright\ rather than \textquotedblleft
PAC-learn.\textquotedblright\ \ Of course, if $\mathsf{P}=\mathsf{NP}$, then
\textit{approximate} learning of Boolean circuits could be done in polynomial
time.}

These results provide a new perspective on one of the oldest problems in
computer science, the \textit{circuit minimization problem}: given a Boolean
circuit $C$, does there exist an equivalent circuit of size at most $s$?
\ Certainly this problem is $\mathsf{NP}$-hard and in $\mathsf{\Sigma}_{2}%
^{p}$. \ Also, by using Bshouty et al.'s algorithm, we can find a circuit
whose size is within an $O\left(  n/\log n\right)  $\ factor of minimal in
$\mathsf{ZPP}^{\mathsf{NP}}$. \ Yet after fifty years of research, almost
nothing else is known about the complexity of this problem. \ For example, is
it $\mathsf{\Sigma}_{2}^{p}$-complete? \ Can we approximate the minimum
circuit size in $\mathsf{ZPP}_{||}^{\mathsf{NP}}$?

What our techniques let us say is the following. \ First, there exists an
oracle $A$ such that minimizing circuits with oracle access to $A$\ is not
even approximable in $\mathsf{BPP}_{||}^{\mathsf{NP}^{A}}$. \ Indeed, any
probabilistic algorithm to distinguish the cases \textquotedblleft$C$\ is
minimal\textquotedblright\ and \textquotedblleft there exists an equivalent
circuit for $C$ of size $s$,\textquotedblright\ using fewer than $s$\ adaptive
$\mathsf{NP}$ queries, would have to use nonrelativizing techniques. \ If one
wished, one could take this as evidence that the true complexity of the
circuit minimization problem should be $\mathsf{P}^{\mathsf{NP}}$ rather than
$\mathsf{P}_{||}^{\mathsf{NP}}$. \ On the other hand, one cannot rule out even
a \textquotedblleft black-box\textquotedblright\ circuit minimization
algorithm (that is, an algorithm that treats $C$ itself as an oracle)\ in
$\mathsf{P}_{||}^{\mathsf{NP}}$,\ without also showing that $\mathsf{P}%
\neq\mathsf{NP}$.

From a learning theory perspective, perhaps what is most interesting about our
results is that they show a clear tradeoff between two complexities: the
complexity of the learner who queries the target function $f$, and the
complexity of the resulting computational problem that the learner has to
solve. \ If the learner is a $\mathsf{ZPP}^{\mathsf{NP}^{f}}$\ machine, then
the problem is easy; if the learner is a $\mathsf{ZPP}_{||}^{\mathsf{NP}^{f}}%
$\ machine, then the problem is (probably) hard; and if the learner is an
$\mathsf{NP}^{f}$\ machine, then there is \textit{no} computational problem
whose solution would suffice to learn $f$.

\subsection{Outlook\label{OUTLOOK}}

\begin{figure}
[ptb]
\begin{center}
\includegraphics[
trim=0.000000in 0.000000in 0.000000in 1.090111in,
height=4.8741in,
width=4.0681in
]%
{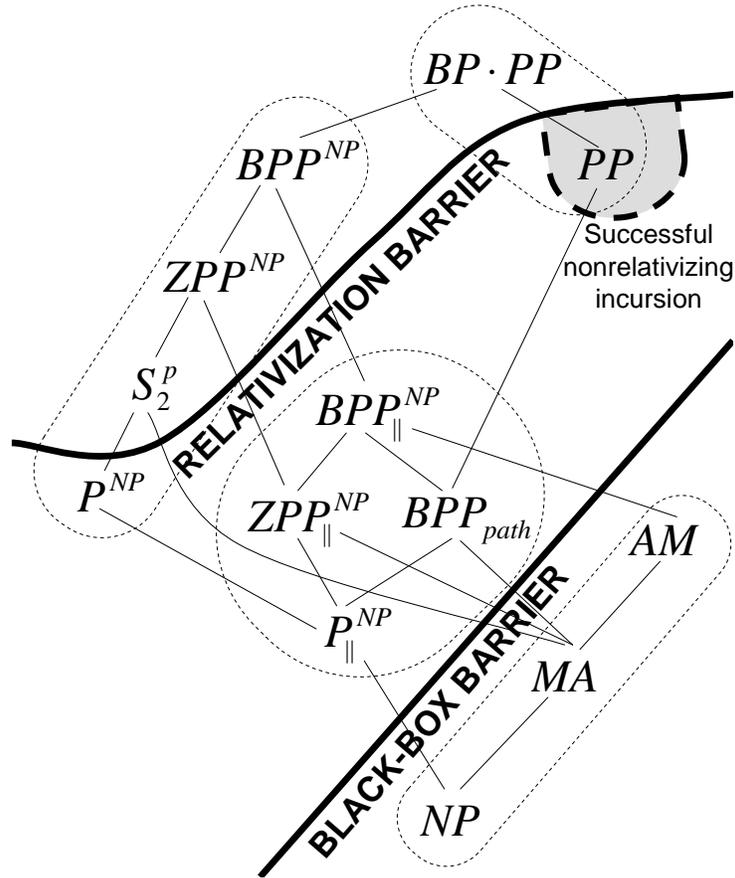}%
\caption{\textquotedblleft Battle map\textquotedblright\ of some complexity
classes between $\mathsf{NP}$\ and $\mathsf{BP\cdot PP}$, in light of this
paper's results. \ Classes that coincide under a plausible derandomization
assumption are grouped together with dashed ovals. \ Below the relativization
barrier, we must use nonrelativizing techniques to show any Karp-Lipton
collapse or superlinear circuit size lower bound. \ Below the black-box
barrier, black-box learning of Boolean circuits is provably impossible.}%
\label{subtlefig}%
\end{center}
\end{figure}

Figure \ref{subtlefig}\ shows the \textquotedblleft battle
map\textquotedblright\ for nonrelativizing circuit lower bounds that emerges
from this paper. \ The figure displays not one but two barriers: a
\textquotedblleft relativization barrier,\textquotedblright\ below which any
Karp-Lipton collapse or superlinear circuit size lower bound will need to use
nonrelativizing techniques; and a \textquotedblleft black-box
barrier,\textquotedblright\ below which black-box learning even of
unrelativized circuits is provably impossible. \ At least for the thirteen
complexity classes shown in the figure, we now know exactly where to draw
these two barriers---something that would have been less than obvious
\textit{a priori}\ (at least to us!).

To switch metaphors, we can think of the barriers as representing
\textquotedblleft phase transitions\textquotedblright\ in the behavior of
complexity classes. \ Below the black-box barrier, we cannot learn circuits
relative to any oracle $A$. \ Between the relativization and black-box
barriers, we can learn Boolean circuits relative to \textit{some} oracles
$A$\ but not others. \ For example, we can learn relative to a
$\mathsf{PSPACE}$\ oracle, since it collapses $\mathsf{P}$\ and $\mathsf{NP}$,
but we cannot learn relative to the oracles in this paper, which
cause\ $\mathsf{PP}$\ and $\mathsf{BPP}_{||}^{\mathsf{NP}}$\ to have
linear-size circuits. \ Finally, above the relativization barrier, we can
learn Boolean circuits relative to \textit{every} oracle $A$.\footnote{There
is one important caveat: in $\mathsf{S}_{2}^{p}$, we currently only know how
to learn self-reducible functions, such as the characteristic functions of
$\mathsf{NP}$-complete problems. \ For if the circuits from the two competing
provers disagree with each other, then we need to know which one to trust.}
\ As we move upward from the black-box barrier toward the relativization
barrier, we can notice \textquotedblleft steam bubbles\textquotedblright%
\ starting to form, as the assumptions needed for black-box learning shift
from implausible ($\mathsf{P}=\mathsf{NP}$), to plausible (the standard
derandomization assumptions that collapse $\mathsf{P}^{\mathsf{NP}}$\ with
$\mathsf{ZPP}^{\mathsf{NP}}$\ and\ $\mathsf{PP}$\ with $\mathsf{BP\cdot PP}$),
and finally to no assumptions at all.

To switch metaphors again, the oracle results have laid before us a rich and
detailed landscape, which a nonrelativizing Lewis-and-Clark expedition might
someday visit more fully.%

\section{\label{PPORACLE}The Oracle for $\mathsf{PP}$}

In this section we construct an oracle relative to which $\mathsf{PP}$\ has
linear-size circuits. \ To do so, we will need a lemma about multilinear
polynomials, which follows from the well-known lower bound of Nisan and
Szegedy \cite{ns} on the approximate degree of the OR\ function.

\begin{lemma}
[Nisan-Szegedy]\label{ns}Let $p:\left\{  0,1\right\}  ^{N}\rightarrow
\mathbb{R}$\ be a real multilinear polynomial of degree at most $\sqrt{N}%
/7$,\ and suppose that $\left\vert p\left(  X\right)  \right\vert \leq\frac
{2}{3}\left\vert p\left(  0^{N}\right)  \right\vert $ for all $X\in\left\{
0,1\right\}  ^{N}$\ with Hamming weight $1$. \ Then there exists an
$X\in\left\{  0,1\right\}  ^{N}$\ such that $\left\vert p\left(  X\right)
\right\vert \geq6\left\vert p\left(  0^{n}\right)  \right\vert $.
\end{lemma}

We now prove the main result.

\begin{theorem}
\label{ppthm}There exists an oracle relative to which $\mathsf{PP}$ has
linear-size circuits.
\end{theorem}

\begin{proof}
For simplicity, we first give an oracle that works for a specific value of
$n$, and then generalize to all $n$ simultaneously. \ Let $M_{1},M_{2},\ldots
$\ be an enumeration of $\mathsf{PTIME}\left(  n^{\log n}\right)  $\ machines.
\ Then it suffices to simulate $M_{1},\ldots,M_{n}$, for in that case every
$M_{i}$\ will be simulated on all but finitely many $n$.

The oracle $A$ will consist of $2^{5n}$\ \textquotedblleft
rows\textquotedblright\ and $n2^{n}$\ \textquotedblleft
columns,\textquotedblright\ with each row labeled by a string $r\in\left\{
0,1\right\}  ^{5n}$, and each column labeled by a pair $\left\langle
i,x\right\rangle $\ where $i\in\left\{  1,\ldots,n\right\}  $\ and
$x\in\left\{  0,1\right\}  ^{n}$. \ Then given a triple $\left\langle
r,i,x\right\rangle $\ as input, $A$ will return the bit $A\left(
r,i,x\right)  $.

We will construct $A$ via an iterative procedure. \ Initially $A$\ is empty
(that is, $A\left(  r,i,x\right)  =0$ for all $r,i,x$). \ Let $A_{t}$\ be the
state of $A$ after the $t^{th}$\ iteration. \ Also, let $M_{i,x}\left(
A\right)  $\ be a Boolean function that equals $1$ if $M_{i}$\ accepts on
input $x\in\left\{  0,1\right\}  ^{n}$ and oracle string $A$, and $0$
otherwise. \ Then to \textit{encode} a row $r$ means to set $A_{t}\left(
r,i,x\right)  :=M_{i,x}\left(  A_{t-1}\right)  $ for all $i,x$. \ At a high
level, our entire procedure will consist of repeating the following two steps,
for all $t\geq1$:

\begin{enumerate}
\item[(1)] Choose a set of rows\ $S\subseteq\left\{  0,1\right\}  ^{5n}$ of
$A_{t-1}$.

\item[(2)] Encode each $r\in S$, and let $A_{t}$ be the result.
\end{enumerate}

The problem, of course, is that each time we encode a row $r$, the
$M_{i,x}\left(  A\right)  $'s might change as a result. \ So we need to show
that, by carefully implementing step (1), we can guarantee that the following
condition holds after a finite number of steps.

\begin{enumerate}
\item[($\mathcal{C}$)] There exists an $r$ such that $A\left(  r,i,x\right)
=M_{i,x}\left(  A\right)  $\ for all $i,x$.
\end{enumerate}

If ($\mathcal{C}$) is satisfied, then clearly $M_{1},\ldots,M_{n}$\ will have
linear-size circuits relative to $A$, since we can just hardwire $r$ into the circuits.

We will use the following fact, which is immediate from the definition of
$\mathsf{PP}$. \ For all $i,x$, there exists a multilinear polynomial
$p_{i,x}\left(  A\right)  $, whose variables are\ the bits of $A$, such that:

\begin{enumerate}
\item[(i)] If $M_{i,x}\left(  A\right)  =1$\ then $p_{i,x}\left(  A\right)
\geq1$.

\item[(ii)] If $M_{i,x}\left(  A\right)  =0$\ then $p_{i,x}\left(  A\right)
\leq-1$.

\item[(iii)] $p_{i,x}$ has degree at most $n^{\log n}$.

\item[(iv)] $\left\vert p_{i,x}\left(  A\right)  \right\vert \leq2^{n^{\log
n}}$ for all $A$.
\end{enumerate}

Now for all integers $0\leq k\leq n^{\log n}$\ and $b\in\left\{  0,1\right\}
$, let%
\[
q_{i,x,b,k}\left(  A\right)  =2^{2k-3}+\left(  2^{k}+\left(  -1\right)
^{b}p_{i,x}\left(  A\right)  \right)  ^{2}.
\]
Then we will use the following polynomial as a progress measure:%
\[
Q\left(  A\right)  =%
{\displaystyle\prod\limits_{i,x}}
{\displaystyle\prod\limits_{b\in\left\{  0,1\right\}  }}
{\displaystyle\prod\limits_{k=0}^{n^{\log n}}}
q_{i,x,b,k}\left(  A\right)  .
\]
Notice that%
\[
\deg\left(  Q\right)  \leq n2^{n}\cdot2\cdot\left(  n^{\log n}+1\right)
\cdot2\deg\left(  p_{i,x}\right)  =2^{n+o\left(  n\right)  }.
\]
Since $1/8\leq q_{i,x,b,k}\left(  A\right)  \leq5\cdot2^{2n^{\log n}}$\ for
all $i,x,b,k$, we also have%
\begin{align*}
Q\left(  A\right)   &  \leq\left(  5\cdot2^{2n^{\log n}}\right)  ^{n2^{n}%
\cdot2\cdot\left(  n^{\log n}+1\right)  }=2^{2^{n+o\left(  n\right)  }},\\
Q\left(  A\right)   &  \geq\left(  \frac{1}{8}\right)  ^{n2^{n}\cdot
2\cdot\left(  n^{\log n}+1\right)  }=2^{-2^{n+o\left(  n\right)  }}%
\end{align*}
for all $A$. \ The key claim is the following.

\textit{At any given iteration, suppose there is no }$r$\textit{ such that, by
encoding }$r$\textit{, we can satisfy condition (}$\mathcal{C}$\textit{).
\ Then there exists a set }$S\subseteq\left\{  0,1\right\}  ^{5n}%
$\textit{\ such that, by encoding each }$r\in S$\textit{, we can increase
}$Q\left(  A\right)  $\textit{\ by at least a factor of }$2$\textit{ (that is,
ensure that }$Q\left(  A_{t}\right)  \geq2Q\left(  A_{t-1}\right)
$\textit{).}

The above claim readily implies that ($\mathcal{C}$) can be satisfied after a
finite number of steps. \ For, by what was said previously, $Q\left(
A\right)  $\ can double at most $2^{n+o\left(  n\right)  }$\ times---and once
$Q\left(  A\right)  $ can no longer double, by assumption\ we can encode an
$r$ that satisfies ($\mathcal{C}$). \ (As a side note, \textquotedblleft
running out of rows\textquotedblright\ is not an issue here, since we can
re-encode rows that were encoded in previous iterations.)

We now prove the claim. \ Call the\ pair $\left\langle i,x\right\rangle $
\textit{sensitive} to row $r$ if encoding $r$\ would change the value of
$M_{i,x}\left(  A\right)  $. \ By hypothesis, for every $r$ there exists an
$\left\langle i,x\right\rangle $\ that is sensitive to $r$. \ So by a counting
argument, there exists a single $\left\langle i,x\right\rangle $\ that is
sensitive to at least $2^{5n}/\left(  n2^{n}\right)  >2^{3n}$\ rows. \ Fix
that $\left\langle i,x\right\rangle $, and let $r_{1},\ldots,r_{2^{3n}}$\ be
the first $2^{3n}$ rows to which $\left\langle i,x\right\rangle $\ is
sensitive. \ Also, given a binary string $Y=y_{1}\ldots y_{2^{3n}}$, let
$S\left(  Y\right)  $\ be the set of all $r_{j}$\ such that $y_{j}=1$, and let
$A^{\left(  Y\right)  }$\ be the oracle obtained by starting from $A$ and then
encoding each $r_{j}\in S\left(  Y\right)  $.

Set $b$ equal to $M_{i,x}\left(  A\right)  $, and set $k$ equal to the least
integer such that $2^{k}\geq\left\vert p_{i,x}\left(  A\right)  \right\vert $.
\ Then we will think of $Q\left(  A\right)  $\ as the product of two
polynomials $q\left(  A\right)  $ and $v\left(  A\right)  $, where $q\left(
A\right)  =q_{i,x,b,k}\left(  A\right)  $,\ and $v\left(  A\right)  =Q\left(
A\right)  /q\left(  A\right)  $ is the product of all other terms in $Q\left(
A\right)  $. \ Notice that $q\left(  A\right)  >0$\ and $v\left(  A\right)
>0$\ for all $A$. \ Also,%
\begin{align*}
q\left(  A\right)   &  =2^{2k-3}+\left(  2^{k}+\left(  -1\right)  ^{b}%
p_{i,x}\left(  A\right)  \right)  ^{2}\\
&  \leq2^{2k-3}+\left(  2^{k}-2^{k-1}\right)  ^{2}\\
&  =\frac{3}{8}\cdot2^{2k}.
\end{align*}
Here the second line follows since $-2^{k}\leq\left(  -1\right)  ^{b}%
p_{i,x}\left(  A\right)  \leq-2^{k-1}$. \ On the other hand, for all
$Y\in\left\{  0,1\right\}  ^{2^{3n}}$ with Hamming weight $1$, we have
$\left(  -1\right)  ^{b}p_{i,x}\left(  A\right)  \geq0$, and therefore%
\begin{align*}
q\left(  A^{\left(  Y\right)  }\right)   &  =2^{2k-3}+\left(  2^{k}+\left(
-1\right)  ^{b}p_{i,x}\left(  A^{\left(  Y\right)  }\right)  \right)  ^{2}\\
&  \geq2^{2k-3}+\left(  2^{k}\right)  ^{2}\\
&  =\frac{9}{8}\cdot2^{2k}\\
&  \geq3q\left(  A\right)  .
\end{align*}

There are now two cases. \ The first is that there exists a $Y$ with Hamming
weight $1$ such that $v\left(  A^{\left(  Y\right)  }\right)  \geq\frac{2}%
{3}v\left(  A\right)  $. \ In this case%
\begin{align*}
Q\left(  A^{\left(  Y\right)  }\right)   &  =q\left(  A^{\left(  Y\right)
}\right)  v\left(  A^{\left(  Y\right)  }\right) \\
&  \geq3q\left(  A\right)  \cdot\frac{2}{3}v\left(  A\right) \\
&  =2q\left(  A\right)  v\left(  A\right) \\
&  =2Q\left(  A\right)  .
\end{align*}
So we simply set $S=S\left(  Y\right)  $\ and are done.

The second case is that $v\left(  A^{\left(  Y\right)  }\right)  <\frac{2}%
{3}v\left(  A\right)  $\ for all $Y$ with Hamming weight $1$. \ In this case,
we can consider $v$ as a real multilinear polynomial in the bits of
$Y\in\left\{  0,1\right\}  ^{2^{3n}}$, of degree at most $\deg\left(
Q\right)  <\sqrt{2^{3n}}/7$. \ Then Lemma \ref{ns}\ implies that there exists
a $Y\in\left\{  0,1\right\}  ^{2^{3n}}$\ such that $\left\vert v\left(
A^{\left(  Y\right)  }\right)  \right\vert =v\left(  A^{\left(  Y\right)
}\right)  \geq6v\left(  A\right)  $. \ Furthermore, for all $Y$ we have%
\[
\frac{q\left(  A^{\left(  Y\right)  }\right)  }{q\left(  A\right)  }\geq
\frac{2^{2k-3}}{\frac{3}{8}\cdot2^{2k}}=\frac{1}{3}.
\]
Hence%
\begin{align*}
Q\left(  A^{\left(  Y\right)  }\right)   &  =q\left(  A^{\left(  Y\right)
}\right)  v\left(  A^{\left(  Y\right)  }\right) \\
&  \geq\frac{1}{3}q\left(  A\right)  \cdot6v\left(  A\right) \\
&  =2q\left(  A\right)  v\left(  A\right) \\
&  =2Q\left(  A\right)  .
\end{align*}
So again we can set $S=S\left(  Y\right)  $. \ This completes the claim.

All that remains is to handle\ $\mathsf{PTIME}\left(  n^{\log n}\right)
$\ machines that could query \textit{any} bit of the oracle string, rather
than just the bits corresponding to a specific $n$. \ The oracle $A$ will now
take as input a \textit{list} of strings $R=\left(  r_{1},\ldots,r_{\ell
}\right)  $, with $r_{\ell}\in\left\{  0,1\right\}  ^{5\cdot2^{\ell}}$\ for
all $\ell$, in addition to $i,x$. \ Call $R$ an $\ell$\textit{-secret}\ if
$A\left(  R,i,x\right)  =M_{i,x}\left(  A\right)  $\ for all $n\leq2^{\ell}$,
$i\in\left\{  1,\ldots,n\right\}  $, and $x\in\left\{  0,1\right\}  ^{n}$.
\ Then we will try to satisfy the following.

\begin{enumerate}
\item[($\mathcal{C}^{\prime}$)] There exists an infinite list of strings
$r_{1}^{\ast},r_{2}^{\ast},\ldots$, ,\ such that $R_{\ell}^{\ast}:=\left(
r_{1}^{\ast},\ldots,r_{\ell}^{\ast}\right)  $ is an $\ell$-secret for all
$\ell\geq1$.
\end{enumerate}

If ($\mathcal{C}^{\prime}$) is satisfied, then clearly each $M_{i}$\ can be
simulated by linear-size circuits. For all $n\geq i$, simply find the smallest
$\ell$\ such that $2^{\ell}\geq n$, then hardwire $R_{\ell}^{\ast}$\ into the
circuit for size $n$.\ \ Since $\ell\leq2n$, this requires at most $5\left(
2^{1}+\cdots+2^{\ell}\right)  \leq20n$ bits.

To construct an oracle $A$ that satisfies ($\mathcal{C}^{\prime}$), we iterate
over all $\ell\geq1$. \ Suppose by induction that $R_{\ell-1}^{\ast}$\ is an
$\left(  \ell-1\right)  $-secret;\ then we want to ensure that $R_{\ell}%
^{\ast}$\ is an $\ell$-secret\ for some $r_{\ell}\in\left\{  0,1\right\}
^{5\cdot2^{\ell}}$. \ To do so, we use a procedure essentially identical to
the one for a specific $n$. \ The only difference is this: previously, all we
needed was a row $r\in\left\{  0,1\right\}  ^{5n}$\ such that no $\left\langle
i,x\right\rangle $\ pair was sensitive to a \textit{particular} change to $r$
(namely, setting $A_{t}\left(  r,i,x\right)  :=M_{i,x}\left(  A_{t-1}\right)
$\ for all $i,x$). \ But in the general case, the \textquotedblleft
row\textquotedblright\ labeled by $R=\left(  r_{1},\ldots,r_{\ell}\right)
$\ consists of all triples $\left\langle R^{\prime},i,x\right\rangle $ such
that $R^{\prime}=\left(  r_{1},\ldots,r_{\ell},r_{\ell+1}^{\prime}%
,\ldots,r_{L}^{\prime}\right)  $\ for some $L\geq\ell$\ and $r_{\ell
+1}^{\prime},\ldots,r_{L}^{\prime}$. \ Furthermore, we do not yet know how
later iterations will affect this \textquotedblleft row.\textquotedblright%
\ \ So we should call a pair $\left\langle i,x\right\rangle $
\textquotedblleft sensitive\textquotedblright\ to $R$, if there is
\textit{any} oracle $A^{\prime}$\ such that (1) $A^{\prime}$ disagrees with
$A$ only in row $R$, and (2) $M_{i,x}\left(  A^{\prime}\right)  \neq
M_{i,x}\left(  A\right)  $.

Fortunately, this new notion of sensitivity requires no significant change to
the proof. \ Suppose that for every row $R$ of the form $\left(  r_{1}^{\ast
},\ldots,r_{\ell-1}^{\ast},r_{\ell}\right)  $ there exists an $\left\langle
i,x\right\rangle $ that is sensitive to $R$. \ Then as before, there exists an
$\left\langle i^{\prime},x^{\prime}\right\rangle $ that is sensitive to at
least $2^{5\cdot2^{\ell}}/\left(  2^{2\ell}2^{2^{\ell}+1}\right)  >2^{3n}%
$\ rows of that form. \ For each of those rows $R$, fix a change to $R$\ to
which $\left\langle i^{\prime},x^{\prime}\right\rangle $\ is sensitive. \ We
thereby obtain a polynomial $Q\left(  A\right)  $\ with the same properties as
before---in particular, there exists a string $Y\in\left\{  0,1\right\}
^{2^{3n}}$\ such that $Q\left(  A^{\left(  Y\right)  }\right)  \geq2Q\left(
A\right)  $.
\end{proof}

Let us make three remarks about Theorem \ref{ppthm}.

\begin{enumerate}
\item[(1)] If we care about constants,\ it is clear that the advice $r$ can be
reduced to $3n+o\left(  n\right)  $\ bits for a specific $n$, or $12n+o\left(
n\right)  $\ for all $n$ simultaneously. \ Presumably these bounds are not tight.

\item[(2)] One can easily extend Theorem \ref{ppthm}\ to give an oracle
relative to which\ $\mathsf{PE}=\mathsf{PTIME}\left(  2^{O\left(  n\right)
}\right)  $ has linear-size circuits, and hence $\mathsf{PEXP}\subset
\mathsf{P/poly}$ by a padding argument.

\item[(3)] Han, Hemaspaandra, and Thierauf \cite{hht}\ showed that
$\mathsf{MA}\subseteq\mathsf{BPP}_{\mathsf{path}}\subseteq\mathsf{PP}$. \ So
in addition to implying the result of Buhrman, Fortnow, and Thierauf that
$\mathsf{MA}$\ has linear-size circuits relative to an oracle, Theorem
\ref{ppthm}\ also yields the new result that $\mathsf{BPP}_{\mathsf{path}}%
$\ has linear-size circuits relative to an oracle.
\end{enumerate}

Another application of our techniques, the construction of relativized worlds
where $\mathsf{P}^{\mathsf{NP}}=\mathsf{PEXP}$ and $\mathsf{\oplus
P}=\mathsf{PEXP}$, is outlined in Appendix \ref{CRUNCH}.

\section{Quantum Circuit Lower Bounds\label{QUANTUM}}

In this section we show, by a nonrelativizing argument, that $\mathsf{PP}%
$\ does \textit{not} have circuits of size $n^{k}$, not even quantum circuits
with quantum advice. \ We first show that $\mathsf{P}^{\mathsf{PP}}$ does not
have quantum circuits of size $n^{k}$, by a direct diagonalization argument.
\ Our argument will use the following lemma of Aaronson \cite{aar:adv}.

\begin{lemma}
[\textquotedblleft Almost As Good As New Lemma\textquotedblright%
]\label{goodasnew}Suppose a two-outcome measurement of a mixed quantum state
$\rho$\ yields outcome $0$\ with probability $1-\varepsilon$. \ Then after the
measurement, we can recover a state $\widetilde{\rho}$\ such that $\left\Vert
\widetilde{\rho}-\rho\right\Vert _{\operatorname*{tr}}\leq\sqrt{\varepsilon}$.
\end{lemma}

(Recall that the trace distance $\left\Vert \rho-\sigma\right\Vert
_{\operatorname*{tr}}$\ between two mixed states $\rho$\ and $\sigma$\ is the
maximum bias with which those states can be distinguished via a single
measurement. \ In particular, trace distance satisfies the triangle inequality.)

\begin{theorem}
\label{ppp}$\mathsf{P}^{\mathsf{PP}}$ does not have quantum circuits of size
$n^{k}$\ for any fixed $k$. \ Furthermore, this holds even if the circuits can
use quantum advice.
\end{theorem}

\begin{proof}
For simplicity, let us first explain why $\mathsf{P}^{\mathsf{PP}}$\ does not
have \textit{classical} circuits of size $n^{k}$. \ Fix an input length $n$,
and let $x_{1},\ldots,x_{2^{n}}$\ be a lexicographic ordering of $n$-bit
strings. \ Also, let $\mathcal{C}$ be the set of all circuits of size $n^{k}$,
and let $\mathcal{C}_{t}\subseteq\mathcal{C}$\ be the subset of circuits in
$\mathcal{C}$ that correctly decide the first $t$ inputs $x_{1},\ldots,x_{t}$.
\ Then we define the language $L\cap\left\{  0,1\right\}  ^{n}$ by the
following iterative procedure. \ First, if at least half of the circuits in
$\mathcal{C}$ accept $x_{1}$, then set $x_{1}\notin L$, and otherwise set
$x_{1}\in L$. \ Next, if at least half of the circuits in $\mathcal{C}_{1}%
$\ accept $x_{2}$, then set $x_{2}\notin L$, and otherwise set $x_{2}\in L$.
\ In general, let $N=\left\lceil \log_{2}\left\vert \mathcal{C}^{\prime
}\right\vert \right\rceil +1$. \ Then for all $t<N$, if at least half of the
circuits in $\mathcal{C}_{t}$\ accept $x_{t+1}$, then set $x_{t+1}\notin L$,
and otherwise set $x_{t+1}\in L$.\ \ Finally, set $x_{t}\notin L$ for all
$t>N$.

It is clear that the resulting language $L$ is in $\mathsf{P}^{\mathsf{PP}}$.
\ Given an input $x_{t}$, we just reject if $t>N$, and otherwise call the
$\mathsf{PP}$\ oracle $t$ times, to decide if $x_{i}\in L$\ for each
$i\in\left\{  1,\ldots,t\right\}  $. \ Note that, once we know $x_{1}%
,\ldots,x_{i}$, we can decide in polynomial time whether a given circuit
belongs to $\mathcal{C}_{i}$, and can therefore decide in $\mathsf{PP}%
$\ whether the majority of circuits in $\mathcal{C}_{i}$\ accept or reject
$x_{i+1}$. \ On the other hand, our construction guarantees that $\left\vert
\mathcal{C}_{t+1}\right\vert \leq\left\vert \mathcal{C}_{t}\right\vert /2$ for
all $t<N$. \ Therefore $\left\vert \mathcal{C}_{N}\right\vert \leq\left\vert
\mathcal{C}\right\vert /2^{N}=1/2$, which means that $\mathcal{C}_{N}$\ is
empty, and hence no circuit in $\mathcal{C}$\ correctly decides $x_{1}%
,\ldots,x_{N}$.

The above argument extends naturally to quantum circuits. \ Let $\mathcal{C}$
be the set of all quantum circuits of size $n^{k}$, over a basis of (say)
Hadamard and Toffoli gates.\footnote{Shi \cite{shi:gate} showed that this
basis is universal. \ Any finite, universal set of gates with rational
amplitudes would work equally well.} \ (Note that these circuits need not be
bounded-error.) \ Then the first step is to amplify each circuit
$C\in\mathcal{C}$ a polynomial number times, so that if $C$'s initial error
probability was at most $1/3$, then its new error probability is at most (say)
$2^{-10n}$. \ Let $\mathcal{C}^{\prime}$\ be the resulting set of amplified
circuits. \ Now let $\left\vert \psi_{0}\right\rangle $ be a uniform
superposition over all descriptions of circuits in $\mathcal{C}^{\prime}$,
together with an \textquotedblleft answer register\textquotedblright\ that is
initially set to $\left\vert 0\right\rangle $:%
\[
\left\vert \psi_{0}\right\rangle :=\frac{1}{\sqrt{\left\vert \mathcal{C}%
^{\prime}\right\vert }}\sum_{C\in\mathcal{C}^{\prime}}\left\vert
C\right\rangle \left\vert 0\right\rangle .
\]
For each input $x_{t}\in\left\{  0,1\right\}  ^{n}$, let $U_{t}$\ be a unitary
transformation that maps $\left\vert C\right\rangle \left\vert 0\right\rangle
$\ to $\left\vert C\right\rangle \left\vert C\left(  x_{t}\right)
\right\rangle $\ for each $C\in\mathcal{C}^{\prime}$,\ where $\left\vert
C\left(  x_{t}\right)  \right\rangle $\ is the output of $C$ on input $x_{t}$.
\ (In general, $\left\vert C\left(  x_{t}\right)  \right\rangle $\ will be a
superposition of $\left\vert 0\right\rangle $\ and $\left\vert 1\right\rangle
$.) \ To implement $U_{t}$, we simply simulate running $C$\ on $x_{t}$, and
then run the simulation in reverse to uncompute garbage qubits.

Let $N=\left\lceil \log_{2}\left\vert \mathcal{C}^{\prime}\right\vert
\right\rceil +2$. \ Also, given an input $x_{t}$, let $L\left(  x_{t}\right)
=1$\ if $x_{t}\in L$\ and $L\left(  x_{t}\right)  =0$\ otherwise. \ Fix
$t<N$,\ and suppose by induction that we have already set $L\left(
x_{i}\right)  $\ for all $i\leq t$. \ Then we will use the following quantum
algorithm, called $\mathcal{A}_{t}$, to set $L\left(  x_{t+1}\right)
$.\medskip

\qquad\texttt{Set }$\left\vert \psi\right\rangle :=\left\vert \psi
_{0}\right\rangle $

\qquad\texttt{For }$i:=1$\texttt{\ to }$t$

\qquad\texttt{\qquad Set }$\left\vert \psi\right\rangle :=U_{i}\left\vert
\psi\right\rangle $\texttt{\ }

\qquad\texttt{\qquad Measure the answer register}

\qquad\texttt{\qquad If the measurement outcome is not }$L\left(
x_{i}\right)  $\texttt{, then FAIL}

\qquad\texttt{Next }$i$

\qquad\texttt{Set }$\left\vert \psi\right\rangle :=U_{t+1}\left\vert
\psi\right\rangle $

\qquad\texttt{Measure the answer register\medskip}

Say that $\mathcal{A}_{t}$\ succeeds if it outputs $L\left(  x_{i}\right)
$\ for all $x_{1},\ldots,x_{t}$. \ \textit{Conditioned} on $\mathcal{A}_{t}$
succeeding, if the final measurement yields the outcome $\left\vert
1\right\rangle $\ with probability at least $1/2$, then set $L\left(
x_{t+1}\right)  :=0$, and otherwise set $L\left(  x_{t+1}\right)  :=1$.
\ Finally, set $L\left(  x_{t}\right)  :=0$\ for all $t>N$.

By a simple extension of the result $\mathsf{BQP}\subseteq\mathsf{PP}$\ due to
Adleman, DeMarrais, and Huang \cite{adh}, Aaronson \cite{aar:pp} showed that
polynomial-time quantum computation with postselected\ measurement can be
simulated in $\mathsf{PP}$ (indeed the two are equivalent; that is,
$\mathsf{PostBQP}=\mathsf{PP}$). \ In particular, a $\mathsf{PP}$\ machine can
simulate the postselected quantum algorithm $\mathcal{A}_{t}$ above,\ and
thereby decide whether the final measurement will yield $\left\vert
0\right\rangle $\ or $\left\vert 1\right\rangle $\ with greater probability,
conditioned on all previous measurements having yielded the correct outcomes.
\ It follows that $L\in\mathsf{P}^{\mathsf{PP}}$.

On the other hand, suppose by way of contradiction that there exists a quantum
circuit $C\in\mathcal{C}^{\prime}$ that outputs $L\left(  x_{t}\right)
$\ with probability at least $1-2^{-10n}$ for all $t$. \ Then the probability
that $C$ succeeds on $x_{1},\ldots,x_{N}$\ \textit{simultaneously} is at least
(say) $0.9$, by Lemma \ref{goodasnew}\ together with the triangle inequality.
\ Hence the probability that $\mathcal{A}_{t}$\ succeeds on $x_{1}%
,\ldots,x_{N}$\ is at least $0.9/\left\vert \mathcal{C}^{\prime}\right\vert $.
\ Yet by construction, $\mathcal{A}_{t}$\ succeeds with probability at most
$1/2^{t}$, which is less than $0.9/\left\vert \mathcal{C}^{\prime}\right\vert
$\ when $t=N-1$. \ This yields the desired contradiction.

Finally, to incorporate quantum advice of size $s=n^{k}$, all we need to do is
add an $s$-qubit \textquotedblleft quantum advice register\textquotedblright%
\ to $\left\vert \psi_{0}\right\rangle $, which $U_{t}$'s can use when
simulating the circuits. \ We initialize this advice register to the maximally
mixed state on $s$\ qubits. \ The key fact (see \cite{aar:adv}\ for
example)\ is that, whatever the \textquotedblleft true\textquotedblright%
\ advice state $\left\vert \phi\right\rangle $, we can decompose the maximally
mixed state into%
\[
\frac{1}{2^{s}}\sum_{j=1}^{2^{s}}\left\vert \phi_{j}\right\rangle \left\langle
\phi_{j}\right\vert ,
\]
where $\left\vert \phi_{1}\right\rangle ,\ldots,\left\vert \phi_{2^{s}%
}\right\rangle $\ form an orthonormal basis and $\left\vert \phi
_{1}\right\rangle =\left\vert \phi\right\rangle $. \ By linearity, we can then
track the evolution of each of these $2^{s}$\ components independently. \ So
the previous argument goes through as before, if we set $N=\left\lceil
\log_{2}\left\vert \mathcal{C}^{\prime}\right\vert \right\rceil +s+2$. \ (Note
that we are assuming the advice states are suitably amplified,\ which
increases the running time of $\mathcal{A}_{t}$\ by at most a polynomial factor.)
\end{proof}

Similarly, for all time-constructible functions $f\left(  n\right)  \leq2^{n}%
$, one can show that the class $\mathsf{DTIME}\left(  f\left(  n\right)
\right)  ^{\mathsf{PP}}$ does not have quantum circuits of size $f\left(
n\right)  /n^{2}$. \ So for example, $\mathsf{E}^{\mathsf{PP}}$\ requires
quantum circuits of exponential size.

Having shown a quantum circuit lower bound for $\mathsf{P}^{\mathsf{PP}}$, we
now bootstrap our way down to $\mathsf{PP}$. \ To do so, we use the following
\textquotedblleft quantum Karp-Lipton theorem.\textquotedblright\ \ Here
$\mathsf{BQP}/\mathsf{poly}$\ is $\mathsf{BQP}$\ with polynomial-size
classical advice, $\mathsf{BQP}/\mathsf{qpoly}$\ is $\mathsf{BQP}$\ with
polynomial-size quantum advice, $\mathsf{QMA}$\ is like $\mathsf{MA}$\ but
with quantum verifiers and quantum witnesses, and $\mathsf{QCMA}$\ is like
$\mathsf{MA}$\ but with quantum verifiers and \textit{classical} witnesses.
\ Also, recall that the counting hierarchy $\mathsf{CH}$\ is the union of
$\mathsf{PP}$, $\mathsf{PP}^{\mathsf{PP}}$, $\mathsf{PP}^{\mathsf{PP}%
^{\mathsf{PP}}}$, and so on.

\begin{theorem}
\label{ipthm}If $\mathsf{PP}\subset\mathsf{BQP}/\mathsf{poly}$\ then
$\mathsf{QCMA}=\mathsf{PP}$, and indeed $\mathsf{CH}$ collapses\ to
$\mathsf{QCMA}$. \ Likewise, if $\mathsf{PP}\subset\mathsf{BQP}/\mathsf{qpoly}%
$\ then $\mathsf{CH}$ collapses to $\mathsf{QMA}$.
\end{theorem}

\begin{proof}
Let $L$\ be a language in $\mathsf{CH}$. \ It is clear that we could decide
$L$ in quantum polynomial time, if we were given polynomial-size quantum
circuits for a $\mathsf{PP}$-complete language such as \textsc{MajSat}. \ For
Fortnow and Rogers \cite{fr}\ showed that $\mathsf{BQP}$\ is \textquotedblleft
low\textquotedblright\ for $\mathsf{PP}$; that is, $\mathsf{PP}^{\mathsf{BQP}%
}=\mathsf{PP}$. \ So we could use the quantum circuits for \textsc{MajSat}\ to
collapse $\mathsf{PP}^{\mathsf{PP}}$ to $\mathsf{PP}^{\mathsf{BQP}%
}=\mathsf{PP}$ to $\mathsf{BQP}$, and similarly for all higher levels of
$\mathsf{CH}$.

Assume $\mathsf{PP}\subset\mathsf{BQP}/\mathsf{poly}$; then clearly
$\mathsf{P}^{\mathsf{\#P}}=\mathsf{P}^{\mathsf{PP}}$ is contained in
$\mathsf{BQP}/\mathsf{poly}$\ as well. \ So in $\mathsf{QCMA}$\ we can do the
following: first guess a bounded-error quantum circuit $C$ for computing the
permanent of a $\operatorname*{poly}\left(  n\right)  \times
\operatorname*{poly}\left(  n\right)  $\ matrix over a finite field
$\mathbb{F}_{p}$, for some prime $p=\Theta\left(  \operatorname*{poly}\left(
n\right)  \right)  $. \ (For convenience, here $\operatorname*{poly}\left(
n\right)  $\ means \textquotedblleft a sufficiently large polynomial depending
on $L$.\textquotedblright) \ Then verify that with $1-o\left(  1\right)  $
probability, $C$ works on at least a $1-1/\operatorname*{poly}\left(
n\right)  $ fraction of matrices. \ To do so, simply simulate the interactive
protocol for the permanent due to Lund, Fortnow, Karloff, and Nisan
\cite{lfkn}, but with $C$ in place of the prover. \ Next, use the random
self-reducibility of the permanent\ to generate a new circuit $C^{\prime}%
$\ that, with $1-o\left(  1\right)  $ probability, is correct on
\textit{every} $\operatorname*{poly}\left(  n\right)  \times
\operatorname*{poly}\left(  n\right)  $\ matrix over $\mathbb{F}_{p}$. \ Since
\textsc{Permanent}\ is $\mathsf{\#P}$-complete over all fields of
characteristic $p\neq2$ \cite{valiant}, we can then use $C^{\prime}$\ to
decide \textsc{MajSat} instances of size $\operatorname*{poly}\left(
n\right)  $, and therefore the language $L$ as well.

The case $\mathsf{PP}\subset\mathsf{BQP}/\mathsf{qpoly}$\ is essentially
identical, except that in $\mathsf{QMA}$\ we guess a quantum circuit with
quantum advice. \ That quantum advice states cannot be reused indefinitely
does not present a problem here: we simply guess a boosted circuit, or else
$\operatorname*{poly}\left(  n\right)  $ copies of the original circuit.
\end{proof}

By combining Theorems \ref{ppp} and \ref{ipthm},\ we immediately obtain the following.

\begin{corollary}
\label{ppnk}$\mathsf{PP}$\ does not have quantum circuits of size $n^{k}$ for
any fixed $k$, not even quantum circuits with quantum advice.
\end{corollary}

\begin{proof}
Suppose by contradiction that $\mathsf{PP}$\ had such circuits. \ Then
certainly $\mathsf{PP}\subset\mathsf{BQP}/\mathsf{qpoly}$, so $\mathsf{QMA}%
=\mathsf{PP}=\mathsf{P}^{\mathsf{PP}}=\mathsf{CH}$\ by Theorem \ref{ipthm}.
\ But $\mathsf{P}^{\mathsf{PP}}$\ does \textit{not} have such circuits by
Theorem \ref{ppp}, and therefore neither does $\mathsf{PP}$.
\end{proof}

More generally, for all $f\left(  n\right)  \leq2^{n}$\ we find that
$\mathsf{PTIME}\left(  f\left(  f\left(  n\right)  \right)  \right)
$\ requires quantum circuits of size approximately $f\left(  n\right)  $.
\ For example, $\mathsf{PEXP}$\ requires quantum circuits of \textquotedblleft
half-exponential\textquotedblright\ size.

Finally, we point out a quantum analogue of Buhrman, Fortnow, and Thierauf's
classical nonrelativizing separation \cite{bft}.

\begin{theorem}
\label{qmaexp}$\mathsf{QCMA}_{\mathsf{EXP}}\not \subset \mathsf{BQP}%
/\mathsf{poly}$, and $\mathsf{QMA}_{\mathsf{EXP}}\not \subset \mathsf{BQP}%
/\mathsf{qpoly}$.
\end{theorem}

\begin{proof}
Suppose by contradiction that $\mathsf{QCMA}_{\mathsf{EXP}}\subset
\mathsf{BQP}/\mathsf{poly}$. \ Then clearly $\mathsf{EXP}\subset
\mathsf{BQP}/\mathsf{poly}$\ as well. \ Babai, Fortnow, and Lund
\cite{bfl}\ showed that any language in $\mathsf{EXP}$\ has a two-prover
interactive protocol where the provers are in $\mathsf{EXP}$. \ We can
simulate such a protocol in $\mathsf{QCMA}$\ as follows:\ first guess
(suitably amplified) $\mathsf{BQP}/\mathsf{poly}$ circuits computing the
provers' strategies.\ \ Then simulate the provers and verifier, and accept if
and only if the verifier accepts. \ It follows that $\mathsf{EXP}%
=\mathsf{QCMA}$, and therefore $\mathsf{QCMA}=\mathsf{P}^{\mathsf{PP}}$ as
well. \ So by padding, $\mathsf{QCMA}_{\mathsf{EXP}}=\mathsf{EXP}%
^{\mathsf{PP}}$. \ But we know from Theorem \ref{ppp} that $\mathsf{EXP}%
^{\mathsf{PP}}\not \subset \mathsf{BQP}/\mathsf{poly}$, which yields the
desired contradiction. \ The proof that $\mathsf{QMA}_{\mathsf{EXP}%
}\not \subset \mathsf{BQP}/\mathsf{qpoly}$\ is essentially identical, except
that we guess quantum circuits with quantum advice.
\end{proof}

One can strengthen Theorem \ref{qmaexp}\ to show that $\mathsf{QMA}%
_{\mathsf{EXP}}$\ requires quantum circuits of half-exponential size.
\ However, in contrast to the case for $\mathsf{PEXP}$, here the bound does
not scale down to $\mathsf{QMA}$. \ Indeed, it turns out that the smallest $f$
for which we get \textit{any} superlinear circuit size lower bound for
$\mathsf{QMATIME}\left(  f\left(  n\right)  \right)  $\ is itself half-exponential.

\section{\label{PARORACLE}The Oracle for $\mathsf{BPP}_{||}^{\mathsf{NP}}$}

In this section we construct an oracle relative to which $\mathsf{BPP}%
_{||}^{\mathsf{NP}}$\ has linear-size circuits.

\begin{theorem}
\label{bppnp}There exists an oracle relative to which $\mathsf{BPP}%
_{||}^{\mathsf{NP}}$\ has linear-size circuits.
\end{theorem}

\begin{proof}
As in Theorem \ref{ppthm}, we first give an oracle $A$ that works for a
specific value of $n$. \ Let $M_{1},M_{2},\ldots$\ be an enumeration of
\textquotedblleft syntactic\textquotedblright\ $\mathsf{BPTIME}\left(  n^{\log
n}\right)  _{||}^{\mathsf{NP}}$\ machines, where syntactic means not
necessarily satisfying the promise. \ Then it suffices to simulate
$M_{1},\ldots,M_{n}$. \ We assume without loss of generality that only the
$\mathsf{NP}$\ oracle\ (not the $M_{i}$'s themselves) query $A$, and that each
$\mathsf{NP}$\ call is actually an $\mathsf{NTIME}\left(  n\right)  $\ call
(so in particular, it involves at most $n^{\log n}$\ queries to $A$). \ Let
$M_{i,x,z}\left(  A\right)  $\ be a Boolean function that equals $1$ if
$M_{i}$\ accepts on input $x\in\left\{  0,1\right\}  ^{n}$, random string
$z\in\left\{  0,1\right\}  ^{n^{\log n}}$, and oracle $A$, and $0$ otherwise.
\ Then let $p_{i,x}\left(  A\right)  :=\operatorname*{EX}_{z}\left[
M_{i,x,z}\left(  A\right)  \right]  $\ be the probability that $M_{i}%
$\ accepts $x$.

The oracle $A$ will consist of $2^{3n}$\ rows\ and $n2^{n}$\ columns,\ with
each row labeled by $r\in\left\{  0,1\right\}  ^{3n}$, and each column labeled
by an $\left\langle i,x\right\rangle $\ pair where $i\in\left\{
1,\ldots,n\right\}  $\ and $x\in\left\{  0,1\right\}  ^{n}$. \ We will
construct $A$ via an iterative procedure $\mathcal{P}$. \ Initially $A$\ is
empty (that is, $A\left(  r,i,x\right)  =0$ for all $r,i,x$).\ \ Let $A_{t}%
$\ be the state of $A$ after the $t^{th}$\ iteration. \ Then to
\textit{encode} a row $r$ means to set $A_{t}\left(  r,i,x\right)
:=\operatorname*{round}\left(  p_{i,x}\left(  A_{t-1}\right)  \right)  $ for
all $i,x$, where $\operatorname*{round}\left(  p\right)  =1$\ if $p\geq
1/2$\ and $\operatorname*{round}\left(  p\right)  =0$\ if $p<1/2$.

Call an $\left\langle i,x\right\rangle $ pair \textit{sensitive} to row $r$,
if encoding $r$ would change $p_{i,x}\left(  A\right)  $\ by at least
$1/6$.\ \ Then $\mathcal{P}$ consists entirely of repeating the following two
steps, for $t=1,2,3\ldots$:

\begin{enumerate}
\item[(1)] If there exists an $r$ to which no $\left\langle i,x\right\rangle $
is sensitive, then encode $r$ and halt.

\item[(2)] Otherwise, by a counting argument, there exists a pair
$\left\langle j,y\right\rangle $\ that is sensitive to at least $N=2^{3n}%
/\left(  n2^{n}\right)  $\ rows, call them\ $r_{1},\ldots,r_{N}$. \ Let
$A^{\left(  k\right)  }$\ be the oracle obtained by starting from $A$ and then
encoding $r_{k}$. \ Choose an integer $k\in\left\{  1,\ldots,N\right\}  $ (we
will specify how later), and set $A_{t}:=A_{t-1}^{\left(  k\right)  }$.
\end{enumerate}

Suppose $\mathcal{P}$ halts after $t$ iterations, and let $r$ be the row
encoded by step (1). \ Then by assumption, $\left\vert p_{i,x}\left(
A_{t}\right)  -p_{i,x}\left(  A_{t-1}\right)  \right\vert <1/6$\ for all
$i,x$. \ So in particular, if $p_{i,x}\left(  A_{t}\right)  \geq2/3$\ then
$p_{i,x}\left(  A_{t-1}\right)  >1/2$\ and therefore $A_{t}\left(
r,i,x\right)  =1$. \ Likewise, if $p_{i,x}\left(  A_{t}\right)  \leq1/3$\ then
$p_{i,x}\left(  A_{t-1}\right)  <1/2$\ and therefore $A_{t}\left(
r,i,x\right)  =0$. \ It follows that any valid\ $\mathsf{BPTIME}\left(
n^{\log n}\right)  _{||}^{\mathsf{NP}}$\ machine in $\left\{  M_{1}%
,\ldots,M_{n}\right\}  $\ has linear-size circuits relative to $A_{t}$---since
we can just hardwire $r\in\left\{  0,1\right\}  ^{2n}$ into the circuits.

It remains only to show that $\mathcal{P}$\ halts after a finite number of
steps, for some choice of $k$'s. \ Given an input $x$, random string $z$, and
oracle $A$, let $S_{i,x,z}\left(  A\right)  $\ be the set of $\mathsf{NP}%
$\ queries made by $M_{i}$\ that accept.\ \ Then we will use%
\[
W\left(  A\right)  :=\sum_{i,x}\operatorname*{EX}_{z}\left[  \left\vert
S_{i,x,z}\left(  A\right)  \right\vert \right]
\]
as our progress measure. \ Since each $M_{i}$\ can query the $\mathsf{NP}%
$\ oracle at most $n^{\log n}$\ times, clearly $0\leq\left\vert S_{i,x,z}%
\left(  A\right)  \right\vert \leq n^{\log n}$\ for all $i,x,z$, and therefore%
\[
0\leq W\left(  A\right)  \leq n2^{n}\cdot n^{\log n}%
\]
for all $A$. \ On the other hand, we claim that whenever step (2) is executed,
if $k\in\left\{  1,\ldots,N\right\}  $ is chosen uniformly at random then%
\[
\operatorname*{EX}_{k}\left[  W\left(  A^{\left(  k\right)  }\right)  \right]
\geq W\left(  A\right)  +\frac{1}{6}-2^{-n+o\left(  n\right)  }.
\]
So in step (2), we should simply choose $k$\ to maximize $W\left(  A^{\left(
k\right)  }\right)  $. \ For we will then have $W\left(  A_{t}\right)
\geq\left(  1/6-2^{-n+o\left(  n\right)  }\right)  t$ for all $t$, from which
it follows that $\mathcal{P}$\ halts\ after at most%
\[
\frac{n2^{n}\cdot n^{\log n}}{1/6-2^{-n+o\left(  n\right)  }}=2^{n+o\left(
n\right)  }%
\]
iterations.

We now prove the claim. \ Observe that for each accepting $\mathsf{NP}$\ query
$q\in S_{i,x,z}\left(  A\right)  $, there are at most $n^{\log n}$\ rows
$r_{k}$\ such that encoding $r_{k}$ would cause $q\notin S_{i,x,z}\left(
A^{\left(  k\right)  }\right)  $. \ For to change $q$'s output from `accept'
to `reject,' we would have to eliminate (say) the lexicographically first
accepting path of the $\mathsf{NP}$\ oracle, and that path can depend on at
most $n^{\log n}$\ rows of $A$. \ Hence by the union bound, for all $i,x,z,A$
we have%
\begin{align*}
\Pr_{k}\left[  S_{i,x,z}\left(  A\right)  \not \subset S_{i,x,z}\left(
A^{\left(  k\right)  }\right)  \right]   &  \leq\sum_{q\in S_{i,x,z}\left(
A\right)  }\Pr_{k}\left[  q\notin S_{i,x,z}\left(  A^{\left(  k\right)
}\right)  \right] \\
&  \leq\left\vert S_{i,x,z}\left(  A\right)  \right\vert \frac{n^{\log n}}%
{N}\\
&  \leq\frac{n^{2\log n}}{2^{3n}/\left(  n2^{n}\right)  }\\
&  =2^{-2n+o\left(  n\right)  }.
\end{align*}
So in particular, for all $i,x,A$,%
\begin{align*}
\operatorname*{EX}_{k,z}\left[  \left\vert S_{i,x,z}\left(  A^{\left(
k\right)  }\right)  \right\vert \right]   &  \geq\left\vert S_{i,x,z}\left(
A\right)  \right\vert \cdot\Pr_{k,z}\left[  \left\vert S_{i,x,z}\left(
A^{\left(  k\right)  }\right)  \right\vert \geq\left\vert S_{i,x,z}\left(
A\right)  \right\vert \right] \\
&  \geq\left\vert S_{i,x,z}\left(  A\right)  \right\vert \left(
1-2^{-2n+o\left(  n\right)  }\right)
\end{align*}

On the other hand, by assumption there exists a pair $\left\langle
j,y\right\rangle $\ that is sensitive to row $r_{k}$\ for every $k\in\left\{
1,\ldots,N\right\}  $. \ Furthermore, given $y$ and $z$, the output
$M_{j,y,z}\left(  A\right)  $\ of $M_{j}$\ is a function of the $\mathsf{NP}$
oracle responses $S_{j,y,z}\left(  A\right)  $, and can change only if
$S_{j,y,z}\left(  A\right)  $\ changes. \ Therefore%
\[
\Pr_{k,z}\left[  S_{j,y,z}\left(  A^{\left(  k\right)  }\right)  \neq
S_{j,y,z}\left(  A\right)  \right]  \geq\Pr_{k,z}\left[  M_{j,y,z}\left(
A^{\left(  k\right)  }\right)  \neq M_{j,y,z}\left(  A\right)  \right]
\geq\frac{1}{6}.
\]
So by the union bound,%
\begin{align*}
\Pr_{k,z}\left[  \left\vert S_{j,y,z}\left(  A^{\left(  k\right)  }\right)
\right\vert >\left\vert S_{j,y,z}\left(  A\right)  \right\vert \right]   &
\geq\Pr_{k,z}\left[  S_{j,y,z}\left(  A^{\left(  k\right)  }\right)  \neq
S_{j,y,z}\left(  A\right)  \right]  -\Pr_{k,z}\left[  S_{j,y,z}\left(
A\right)  \not \subset S_{j,y,z}\left(  A^{\left(  k\right)  }\right)  \right]
\\
&  \geq\frac{1}{6}-2^{-2n+o\left(  n\right)  }.
\end{align*}
Putting it all together,%
\begin{align*}
\operatorname*{EX}_{k}\left[  W\left(  A^{\left(  k\right)  }\right)  \right]
&  =\sum_{i,x}\operatorname*{EX}_{k,z}\left[  \left\vert Q_{i,x,z}\left(
A^{\left(  k\right)  }\right)  \right\vert \right] \\
&  \geq\frac{1}{6}-2^{-2n+o\left(  n\right)  }+\sum_{i,x}\left\vert
S_{i,x,z}\left(  A\right)  \right\vert \left(  1-2^{-2n+o\left(  n\right)
}\right) \\
&  =\frac{1}{6}-2^{-2n+o\left(  n\right)  }+\left(  1-2^{-2n+o\left(
n\right)  }\right)  W\left(  A\right) \\
&  =W\left(  A\right)  +\frac{1}{6}-2^{-n+o\left(  n\right)  },
\end{align*}
which completes the claim.

To handle all values of $n$ simultaneously, we use exactly the same trick as
in Theorem \ref{ppthm}. \ That is, we replace $r$ by an $\ell$-tuple
$R=\left(  r_{1},\ldots,r_{\ell}\right)  $ where $r_{\ell}\in\left\{
0,1\right\}  ^{3\cdot2^{\ell}}$; define the \textquotedblleft
row\textquotedblright\ $\mathcal{R}_{\ell}$\ to consist of all triples
$\left\langle R_{L}^{\prime},i,x\right\rangle $ such that $L\geq\ell$\ and
$r_{h}^{\prime}=r_{h}$\ for all $h\leq\ell$; and call the pair $\left\langle
i,x\right\rangle $ \textquotedblleft sensitive\textquotedblright\ to row
$\mathcal{R}_{\ell}$ if there is any oracle $A^{\prime}$\ that disagrees with
$A$ only in $\mathcal{R}_{\ell}$, such that $\left\vert p_{i,x}\left(
A^{\prime}\right)  -p_{i,x}\left(  A\right)  \right\vert \geq1/6$. \ We then
run the procedure $\mathcal{P}$\ repeatedly to encode $r_{1},r_{2},\ldots$,
where \textquotedblleft encoding\textquotedblright\ $r_{\ell}$\ means setting
$A_{t}\left(  R_{\ell},i,x\right)  :=\operatorname*{round}\left(
p_{i,x}\left(  A_{t-1}\right)  \right)  $\ for all $n\leq2^{\ell}$%
,\ $i\in\left\{  1,\ldots,n\right\}  $, and $x\in\left\{  0,1\right\}  ^{n}$.
\ The rest of the proof goes through as before.
\end{proof}

Let us make six remarks about Theorem \ref{bppnp}.

\begin{enumerate}
\item[(1)] An immediate corollary is that any Karp-Lipton collapse to
$\mathsf{BPP}_{||}^{\mathsf{NP}}$ would require nonrelativizing techniques.
\ For relative to the oracle $A$ from the theorem, we have $\mathsf{NP}%
\subseteq\mathsf{BPP}_{||}^{\mathsf{NP}}\subset\mathsf{P/poly}$. \ On the
other hand, if\ $\mathsf{PH}^{A}=\mathsf{BPP}_{||}^{\mathsf{NP}^{A}}$, then
$\mathsf{BPP}_{||}^{\mathsf{NP}^{A}}$\ would not have linear-size circuits by
Kannan's Theorem \cite{kannan} (which relativizes), thereby yielding a contradiction.

\item[(2)] If we care about constants,\ we can reduce the advice $r$ to
$2n+o\left(  n\right)  $\ bits for a specific $n$, or $8n+o\left(  n\right)
$\ for all $n$ simultaneously.

\item[(3)] As with Theorem \ref{ppthm},\ one can easily modify Theorem
\ref{bppnp}\ to give a relativized world where $\mathsf{BPEXP}_{||}%
^{\mathsf{NP}}\subset\mathsf{P/poly}$. \ Thus, Theorem \ref{bppnp}\ provides
an alternate generalization of the result of Buhrman, Fortnow, and Thierauf
\cite{bft}\ that $\mathsf{MA}_{\mathsf{EXP}}\subset\mathsf{P/poly}$\ relative
to an oracle.

\item[(4)] Since $\mathsf{BPP}_{\mathsf{path}}\subseteq\mathsf{BPP}%
_{||}^{\mathsf{NP}}$ (as is not hard to show using approximate counting),
Theorem \ref{bppnp} also provides an alternate proof that $\mathsf{BPP}%
_{\mathsf{path}}$\ has linear-size circuits relative to an oracle.

\item[(5)] Completely analogously to Theorem \ref{trifecta}, one can modify
Theorem \ref{bppnp}\ to give oracles relative to which $\mathsf{P}%
^{\mathsf{NP}}=\mathsf{BPEXP}_{||}^{\mathsf{NP}}$\ and $\mathsf{\oplus
P}=\mathsf{BPEXP}_{||}^{\mathsf{NP}}$.

\item[(6)] For any function $f$, the construction of Theorem \ref{bppnp}%
\ actually yields an oracle relative to which $\mathsf{BPP}^{\mathsf{NP}%
\left[  f\left(  n\right)  \right]  }$\ (that is, $\mathsf{BPP}$\ with
$f\left(  n\right)  $\ adaptive $\mathsf{NP}$\ queries) has circuits of size
$O\left(  n+f\left(  n\right)  \right)  $. \ For clearly we can simulate
$f\left(  n\right)  $\ adaptive queries using $2^{f\left(  n\right)  }%
$\ nonadaptive queries. \ We then repeat Theorem \ref{bppnp}\ with the bound
$0\leq W\left(  A\right)  \leq n2^{n}\cdot2^{f\left(  n\right)  }$.
\end{enumerate}

\section{Black-Box Learning in Algorithmica\label{ALG}}

\textquotedblleft Algorithmica\textquotedblright\ is one of Impagliazzo's five
possible worlds \cite{impagliazzo}, the world in which $\mathsf{P}%
=\mathsf{NP}$. \ In this section we show that in Algorithmica, black-box
learning of Boolean circuits is possible in $\mathsf{P}_{||}^{\mathsf{NP}}$.
\ Let us first define what we mean by black-box learning.

\begin{definition}
\label{bblearn}Say that black-box learning is possible in a complexity class
$\mathcal{C}$ if the following holds. \ There exists a $\mathcal{C}$\ machine
$M$ such that, for all Boolean functions $f:\left\{  0,1\right\}
^{n}\rightarrow\left\{  0,1\right\}  $\ with\ circuit complexity at most
$s\left(  n\right)  $, the machine $M^{f}$\ outputs a circuit for $f$ given
$\left\langle 0^{n},0^{s\left(  n\right)  }\right\rangle $\ as input. \ Also,
$M$ has \textit{approximation ratio} $\alpha\left(  n\right)  $\ if for all
$f$, any circuit output by $M$ has size at most $s\left(  n\right)
\alpha\left(  n\right)  $.
\end{definition}

The above definition\ is admittedly somewhat vague, but for most natural
complexity classes $\mathcal{C}$\ it is clear how to make it precise. Firstly,
by \textquotedblleft$\mathcal{C}$\ machine\textquotedblright\ we really mean
\textquotedblleft$\mathcal{FC}$ machine,\textquotedblright\ where
$\mathcal{FC}$\ is the function version of $\mathcal{C}$. \ Secondly, for
semantic classes, we do not care if the machine violates the promise on inputs
not of the form $\left\langle 0^{n},0^{s\left(  n\right)  }\right\rangle
$,\ or oracles $f$\ that do not have circuit complexity at most $s\left(
n\right)  $. \ Let us give a few examples.

\begin{itemize}
\item Almost by definition, black-box learning is possible in $\mathsf{\Sigma
}_{2}^{p}$\ with approximation ratio $1$.

\item As pointed out by Umans \cite{umans:thesis}, the result of Bshouty et
al. \cite{bshouty}\ implies that black-box learning is possible in
$\mathsf{ZPP}^{\mathsf{NP}}$, with approximation ratio $O\left(  n/\log
n\right)  $.

\item Under standard derandomization assumptions, black-box learning is
possible in $\mathsf{P}^{\mathsf{NP}}$\ with approximation ratio $O\left(
n/\log n\right)  $, and in $\mathsf{PP}$\ with approximation ratio $1$. \ For
not only do these assumptions imply that $\mathsf{ZPP}^{\mathsf{NP}%
}=\mathsf{P}^{\mathsf{NP}}$\ and that $\mathsf{BP\cdot PP}=\mathsf{PP}$, but
they also yield a \textit{black-box simulation} of a $\mathsf{ZPP}%
^{\mathsf{NP}}$\ or $\mathsf{BP\cdot PP}$\ algorithm that learns a circuit for
$f$ by just querying an existing circuit $C$ on various inputs (without
\textquotedblleft cheating\textquotedblright\ and looking at $C$).
\end{itemize}

On the other hand:

\begin{proposition}
\label{nomip}Black-box learning is impossible in $\mathsf{NP}$, or for that
matter in $\mathsf{AM}$, $\mathsf{IP}$, or $\mathsf{MIP}$.
\end{proposition}

\begin{proof}
Suppose there are two possibilities: either $f$ is the identically zero
function, or else $f$ is a point function (that is, there exists a $y$ such
that $f\left(  x\right)  =1$\ if and only if $x=y$). \ In both cases $s\left(
n\right)  =O\left(  n\right)  $. \ But since the verifier has only oracle
access to $f$, it is obvious that no polynomially-bounded sequence of messages
from the prover(s) could convince the verifier that $f$ is identically zero.
\ We omit the details, which were worked out by Fortnow and Sipser \cite{fs}.
\end{proof}

We now prove the main result.

\begin{theorem}
\label{peqnp}If $\mathsf{P}=\mathsf{NP}$, then black-box learning is possible
in $\mathsf{P}_{||}^{\mathsf{NP}}$\ (indeed, with approximation ratio $1$.)
\end{theorem}

\begin{proof}
We use a procedure inspired by that of Bshouty et al. \cite{bshouty}.

Fix $n$, and suppose $f:\left\{  0,1\right\}  ^{n}\rightarrow\left\{
0,1\right\}  $\ has circuits of size $s=s\left(  n\right)  $. \ Let
$\mathcal{B}$\ be the set of all circuits of size $s$, so that $\left\vert
\mathcal{B}\right\vert =s^{O\left(  s\right)  }$. \ Also, say that a circuit
$C\in\mathcal{B}$\ \textit{succeeds} on input $x\in\left\{  0,1\right\}  ^{n}$
if $C\left(  x\right)  =f\left(  x\right)  $,\ and \textit{fails} otherwise.
\ Then given a list of inputs $X=\left(  x_{1},x_{2},\ldots\right)  $, let
$\mathcal{B}\left(  X\right)  $\ be the set of circuits in $\mathcal{B}$\ that
succeed on every $x\in X$.

For the remainder of the proof, let $X_{t}=\left(  x_{1},\ldots,x_{t}\right)
$ be a list of $t$ inputs, and for all $0\leq i<t$, let $X_{i}=\left(
x_{1},\ldots,x_{i}\right)  $\ be the prefix of $X_{t}$\ consisting of the
first $i$ inputs\ (so in particular, $X_{0}$ is the empty list). \ Then our
first claim is that there exists an $\mathsf{NP}^{f}$ machine $Q_{t}$ with the
following behavior:

\begin{itemize}
\item If there exists an $X_{t}$\ such that $\left\vert \mathcal{B}\left(
X_{i}\right)  \right\vert \leq\frac{2}{3}\left\vert \mathcal{B}\left(
X_{i-1}\right)  \right\vert $ for all $i\in\left\{  1,\ldots,t\right\}
$,\ then $Q_{t}$ accepts.

\item If for all $X_{t}$\ there exists an $i\in\left\{  1,\ldots,t\right\}
$\ such that $\left\vert \mathcal{B}\left(  X_{i}\right)  \right\vert
\geq\frac{3}{4}\left\vert \mathcal{B}\left(  X_{i-1}\right)  \right\vert $,
then $Q_{t}$ rejects.
\end{itemize}

(As usual, if neither of the two stated conditions hold, then the machine can
behave arbitrarily.)

In what follows, we can assume without loss of generality that $t$\ is
polynomially bounded. \ For, since \textit{some} circuit $C\in\mathcal{B}%
$\ succeeds on every input, we must have $\left\vert \mathcal{B}\left(
X_{i}\right)  \right\vert \geq1$\ for all $i$. \ Therefore $Q_{t}$\ can
accept\ only if $\left\vert \mathcal{B}\right\vert \left(  3/4\right)
^{t}\geq1$, or equivalently if $t=O\left(  s\log s\right)  $.

Let $f\left(  X_{t}\right)  :=\left(  f\left(  x_{1}\right)  ,\ldots,f\left(
x_{t}\right)  \right)  $,\ and let $z$ be a \textquotedblleft witness
string\textquotedblright\ consisting of $X_{t}$ and $f\left(  X_{t}\right)  $.
\ Then given $z$ and $i\leq t$, we can easily decide whether a circuit $C$
belongs to the set $\mathcal{B}\left(  X_{i}\right)  $: we simply check
whether $C\left(  x_{j}\right)  =f\left(  x_{j}\right)  $\ for all $j\leq i$.
\ So by standard results on approximate counting due to Stockmeyer
\cite{stockmeyer} and Sipser \cite{sipser:bpp}, we can approximate the
cardinality $\left\vert \mathcal{B}\left(  X_{i}\right)  \right\vert $\ in
$\mathsf{BPP}^{\mathsf{NP}}$. \ More precisely, for all $t,i$\ there exists a
$\mathsf{P{}romiseBPP}^{\mathsf{NP}}$\ machine $M_{t,i}$\ such that for all
$z=\left\langle X_{t},f\left(  X_{t}\right)  \right\rangle $:

\begin{itemize}
\item If $\left\vert \mathcal{B}\left(  X_{i}\right)  \right\vert \leq\frac
{2}{3}\left\vert \mathcal{B}\left(  X_{i-1}\right)  \right\vert $ then
$M_{t,i}\left(  z\right)  $\ accepts with probability at least $2/3$ (where
the probability is over $M_{t,i}$'s internal randomness).

\item If\ $\left\vert \mathcal{B}\left(  X_{i}\right)  \right\vert \geq
\frac{3}{4}\left\vert \mathcal{B}\left(  X_{i-1}\right)  \right\vert $ then
$M_{t,i}\left(  z\right)  $\ rejects\ with probability least $2/3$.
\end{itemize}

Now by the Sipser-Lautemann Theorem \cite{sipser:bpp,lautemann}, the
assumption $\mathsf{P}=\mathsf{NP}$ implies that $\mathsf{P{}romiseP}%
=\mathsf{P{}romiseBPP}^{\mathsf{NP}}$ as well. \ So we can convert $M_{t,i}%
$\ into a deterministic polynomial-time machine $M_{t,i}^{\prime}$\ such that
for all $z$:

\begin{itemize}
\item If $\left\vert \mathcal{B}\left(  X_{i}\right)  \right\vert \leq\frac
{2}{3}\left\vert \mathcal{B}\left(  X_{i-1}\right)  \right\vert $\ then
$M_{t,i}^{\prime}\left(  z\right)  $ accepts.

\item If $\left\vert \mathcal{B}\left(  X_{i}\right)  \right\vert \geq\frac
{3}{4}\left\vert \mathcal{B}\left(  X_{i-1}\right)  \right\vert $\ then
$M_{t,i}^{\prime}\left(  z\right)  $ rejects.
\end{itemize}

Using $M_{t,i}^{\prime}$, we can then rewrite $Q_{t}$\ as follows.

\textquotedblleft Does there exist a witness $z$, of the form $\left\langle
X_{t},f\left(  X_{t}\right)  \right\rangle $, such that $M_{t,1}^{\prime
}\left(  z\right)  \wedge\cdots\wedge M_{t,t}^{\prime}\left(  z\right)
$?\textquotedblright

This proves the claim, since the above query is clearly in $\mathsf{NP}^{f}$.

To complete the theorem, we will need one other predicate $A_{t}\left(
z,x\right)  $, with the following behavior. \ For all $z=\left\langle
X_{t},f\left(  X_{t}\right)  \right\rangle $\ and $x\in\left\{  0,1\right\}
^{n}$:

\begin{itemize}
\item If $\Pr_{C\in\mathcal{B}\left(  X_{t}\right)  }\left[  C\left(
x\right)  =1\right]  \geq2/3$\ then $A_{t}\left(  z,x\right)  $\ accepts.

\item If $\Pr_{C\in\mathcal{B}\left(  X_{t}\right)  }\left[  C\left(
x\right)  =0\right]  \geq2/3$\ then $A_{t}\left(  z,x\right)  $\ rejects.
\end{itemize}

It is clear that we can implement $A_{t}$\ in $\mathsf{P{}romiseBPP}%
^{\mathsf{NP}}$, again because of approximate counting and the ease of
deciding membership in $\mathcal{B}\left(  X_{t}\right)  $. \ So by the
assumption $\mathsf{P}=\mathsf{NP}$, we can also implement $A_{t}$\ in
$\mathsf{P}$.

Now let $C_{t,z}$\ be the lexicographically first circuit $C\in\mathcal{B}%
$\ such that $C\left(  x\right)  =A_{t}\left(  z,x\right)  $\ for all
$x\in\left\{  0,1\right\}  ^{n}$. \ Notice that $A_{t}\left(  z,x\right)
$\ is an \textit{explicit procedure}: that is, we can evaluate it without
recourse to the oracle for $f$. \ So given $z$, we can find $C_{t,z}$\ in
$\mathsf{\Delta}_{3}^{p}=\mathsf{P}^{\mathsf{NP}^{\mathsf{NP}}}$, and hence
also in $\mathsf{P}$.

Let $t^{\ast}$\ be the maximum $t$\ for which $Q_{t}$\ accepts,\ and let
$z=\left\langle X_{t^{\ast}},f\left(  X_{t^{\ast}}\right)  \right\rangle $\ be
any accepting witness for $Q_{t^{\ast}}$. \ Then for all $x\in\left\{
0,1\right\}  ^{n}$, we have%
\[
\Pr_{C\in\mathcal{B}\left(  X_{t^{\ast}}\right)  }\left[  C\left(  x\right)
=f\left(  x\right)  \right]  \geq\frac{2}{3}.
\]
For otherwise the sequence $\left(  x_{1},\ldots,x_{t^{\ast}},x\right)
$\ would satisfy $Q_{t^{\ast}+1}$, thereby contradicting the maximality of
$t^{\ast}$. \ An immediate corollary is that $A_{t^{\ast}}\left(  z,x\right)
=f\left(  x\right)  $\ for all $x\in\left\{  0,1\right\}  ^{n}$. \ Hence
$C_{t^{\ast},z}$\ is the lexicographically first circuit for $f$,
independently of the particular accepting witness $z$.

The $\mathsf{P}_{||}^{\mathsf{NP}^{f}}$\ learning algorithm now follows
easily. \ For all $t=O\left(  s\log s\right)  $, the algorithm submits the
query $Q_{t}$ to the $\mathsf{NP}$\ oracle. \ It also submits the following
query, called $R_{t,j}$, for all $t=O\left(  s\log s\right)  $\ and
$j=O\left(  s\log s\right)  $:

\textquotedblleft Does there exist a witness $z=\left\langle X_{t},f\left(
X_{t}\right)  \right\rangle $\ satisfying $Q_{t}$, such that the $j^{th}$\ bit
in the description of $C_{t,z}$ is a $1$?\textquotedblright

Using the responses to the $Q_{t}$'s, the algorithm then determines $t^{\ast}%
$. \ Finally it reads a description of $C_{t^{\ast},z}$ off the responses to
the $R_{t^{\ast},j}$'s.
\end{proof}

Theorem \ref{peqnp} has the following easy corollaries. \ First, we cannot
show that a Karp-Lipton collapse to $\mathsf{P}_{||}^{\mathsf{NP}}$\ would
require non-black-box techniques, without also showing $\mathsf{P}%
\neq\mathsf{NP}$. \ Second, if $\mathsf{P}=\mathsf{NP}$, then black-box
learning is possible in $\mathsf{NP/l{}og}$. \ For since the $\mathsf{P}%
_{||}^{\mathsf{NP}}$\ algorithm of Theorem \ref{peqnp} does not take any
input, we simply count how many of its $\mathsf{NP}$\ queries return a
positive answer, and then feed that number as advice to the $\mathsf{NP/l{}%
og}$\ machine.

\section{Open Problems\label{OPEN}}

The main open problem is, of course, to prove better nonrelativizing lower
bounds. \ For example, can we show that $\mathsf{BPP}_{||}^{\mathsf{NP}}%
$\ does not have linear-size circuits? \ To do so, we would presumably need a
nonrelativizing technique that applies directly to the polynomial hierarchy,
without requiring the full strength of $\mathsf{\#P}$. \ Arora, Impagliazzo,
and Vazirani \cite{aiv}\ argue that \textquotedblleft local
checkability,\textquotedblright\ as used for example in the PCP Theorem,
constitutes such a technique (though see Fortnow \cite{fortnow:rel}\ for a
contrary view). \ For us, the relevant question now is not which techniques
are \textquotedblleft truly\textquotedblright\ nonrelativizing, but simply
which ones lead to lower bounds!

Here are a few other problems.

\begin{itemize}
\item Can we show that $\mathsf{P}^{\mathsf{NP}}\neq\mathsf{PEXP}$? \ If so,
then we would obtain perhaps the first nonrelativizing separation of uniform
complexity classes that does not follow immediately from a collapse such as
$\mathsf{IP}=\mathsf{PSPACE}$\ or $\mathsf{MIP}=\mathsf{NEXP}$.

\item Can we show that $\mathsf{PEXP}$\ requires circuits of exponential size,
rather than just half-exponential?

\item As mentioned in Section \ref{PARINTRO}, Bshouty et al.'s algorithm does
not find a \textit{minimal} circuit for a Boolean function $f$, but only a
circuit within an $O\left(  n/\log n\right)  $ factor of
minimal.\footnote{Actually, the algorithm as we stated it gives an $O\left(
n\right)  $\ approximation ratio, but we can improve it to $O\left(  n/\log
n\right)  $\ by replacing \textquotedblleft at least a $1/3$%
\ fraction\textquotedblright\ by \textquotedblleft at least a
$1/\operatorname*{poly}\left(  n\right)  $\ fraction.\textquotedblright} \ Can
we improve this approximation ratio, or alternatively, show that doing so
would require nonrelativizing techniques?

\item Is black-box learning possible in $\mathsf{P}_{||}^{\mathsf{NP}}$\ or
$\mathsf{ZPP}_{||}^{\mathsf{NP}}$, under some computational assumption that we
actually believe (for example, a derandomization assumption)? \ Alternatively,
can we show that black-box learning is \textit{impossible} in $\mathsf{P}%
_{||}^{\mathsf{NP}}$ under some plausible computational assumption?
\end{itemize}

\section{Acknowledgments}

I am grateful to Lance Fortnow for telling me the problem of whether
$\mathsf{PP}$\ has linear-size circuits relative to an oracle, and for
pointing out the implications of my oracle construction for perceptrons and
for the relativized collapse of $\mathsf{PEXP}$. \ I also thank Avi Wigderson
for sponsoring the postdoc during which this work was done and for many
enlightening conversations; and Boaz Barak, Sasha Razborov, Luca Trevisan,
Chris Umans, Umesh Vazirani, Hoeteck Wee, and Chris Wilson for helpful
discussions and correspondence.

\bibliographystyle{plain}
\bibliography{thesis}

\begin{thebibliography}{10}

\bibitem{aar:adv}
S.~Aaronson.
\newblock Limitations of quantum advice and one-way communication.
\newblock {\em Theory of Computing}, 2004.
\newblock To appear. Conference version in \textit{Proc. IEEE Complexity 2004},
  pp. 320-332. quant-ph/0402095.

\bibitem{aar:pp}
S.~Aaronson.
\newblock Quantum computing, postselection, and probabilistic polynomial-time.
\newblock Submitted. quant-ph/0412187, 2004.

\bibitem{adh}
L.~Adleman, J.~DeMarrais, and M.-D. Huang.
\newblock Quantum computability.
\newblock {\em SIAM J. Comput.}, 26(5):1524--1540, 1997.

\bibitem{aiv}
S.~Arora, R.~Impagliazzo, and U.~Vazirani.
\newblock Relativizing versus nonrelativizing techniques: the role of local
  checkability.
\newblock Manuscript, 1992.

\bibitem{bfl}
L.~Babai, L.~Fortnow, and C.~Lund.
\newblock Nondeterministic exponential time has two-prover interactive
  protocols.
\newblock {\em Computational Complexity}, 1(1):3--40, 1991.

\bibitem{bgs}
T.~Baker, J.~Gill, and R.~Solovay.
\newblock Relativizations of the {P=?NP} question.
\newblock {\em SIAM J. Comput.}, 4:431--442, 1975.

\bibitem{beigel}
R.~Beigel.
\newblock Perceptrons, {PP}, and the polynomial hierarchy.
\newblock {\em Computational Complexity}, 4:339--349, 1994.

\bibitem{bshouty}
N.~H. Bshouty, R.~Cleve, R.~Gavald\`{a}, S.~Kannan, and C.~Tamon.
\newblock Oracles and queries that are sufficient for exact learning.
\newblock {\em J. Comput. Sys. Sci.}, 52(3):421--433, 1996.

\bibitem{bfft}
H.~Buhrman, S.~Fenner, L.~Fortnow, and L.~Torenvliet.
\newblock Two oracles that force a big crunch.
\newblock {\em Computational Complexity}, 10(2):93--116, 2001.

\bibitem{bft}
H.~Buhrman, L.~Fortnow, and T.~Thierauf.
\newblock Nonrelativizing separations.
\newblock In {\em Proc. IEEE Conference on Computational Complexity}, pages
  8--12, 1998.

\bibitem{cai}
J.-Y. Cai.
\newblock {$S_{2}^{p}\subseteq ZPP^{NP}$}.
\newblock In {\em Proc. IEEE FOCS}, pages 620--629, 2001.

\bibitem{fortnow:rel}
L.~Fortnow.
\newblock The role of relativization in complexity theory.
\newblock {\em Bulletin of the EATCS}, 52:229--244, February 1994.

\bibitem{fk}
L.~Fortnow and A.~Klivans.
\newblock {NP} with small advice.
\newblock In {\em Proc. IEEE Conference on Computational Complexity}, 2005.
\newblock To appear.

\bibitem{fr}
L.~Fortnow and J.~Rogers.
\newblock Complexity limitations on quantum computation.
\newblock {\em J. Comput. Sys. Sci.}, 59(2):240--252, 1999.
\newblock cs.CC/9811023.

\bibitem{fs}
L.~Fortnow and M.~Sipser.
\newblock Are there interactive protocols for co-{NP} languages?
\newblock {\em Inform. Proc. Lett.}, 28:249--251, 1988.

\bibitem{hht}
Y.~Han, L.~Hemaspaandra, and T.~Thierauf.
\newblock Threshold computation and cryptographic security.
\newblock {\em SIAM J. Comput.}, 26(1):59--78, 1997.

\bibitem{impagliazzo}
R.~Impagliazzo.
\newblock A personal view of average-case complexity.
\newblock In {\em Proc. IEEE Conference on Computational Complexity}, pages
  134--147, 1995.

\bibitem{kannan}
R.~Kannan.
\newblock Circuit-size lower bounds and non-reducibility to sparse sets.
\newblock {\em Information and Control}, 55:40--56, 1982.

\bibitem{kl}
R.~M. Karp and R.~J. Lipton.
\newblock Turing machines that take advice.
\newblock {\em Enseign. Math.}, 28:191--201, 1982.

\bibitem{kobler}
J.~K\"{o}bler and O.~Watanabe.
\newblock New collapse consequences of {NP} having small circuits.
\newblock {\em SIAM J. Comput.}, 28(1):311--324, 1998.

\bibitem{lautemann}
C.~Lautemann.
\newblock {BPP} and the polynomial hierarchy.
\newblock {\em Inform. Proc. Lett.}, 17:215--217, 1983.

\bibitem{lfkn}
C.~Lund, L.~Fortnow, H.~Karloff, and N.~Nisan.
\newblock Algebraic methods for interactive proof systems.
\newblock {\em J. ACM}, 39:859--868, 1992.

\bibitem{mvw}
P.~B. Miltersen, N.~V. Vinodchandran, and O.~Watanabe.
\newblock Super-polynomial versus half-exponential circuit size in the
  exponential hierarchy.
\newblock In {\em COCOON}, pages 210--220, 1999.

\bibitem{mp}
M.~Minsky and S.~Papert.
\newblock {\em Perceptrons (2nd edition)}.
\newblock MIT Press, 1988.
\newblock First appeared in 1968.

\bibitem{ns}
N.~Nisan and M.~Szegedy.
\newblock On the degree of {B}oolean functions as real polynomials.
\newblock {\em Computational Complexity}, 4(4):301--313, 1994.

\bibitem{ny}
H.~Nishimura and T.~Yamakami.
\newblock Polynomial time quantum computation with advice.
\newblock {\em Inform. Proc. Lett.}, 90:195--204, 2003.
\newblock ECCC TR03-059, quant-ph/0305100.

\bibitem{rr}
A.~A. Razborov and S.~Rudich.
\newblock Natural proofs.
\newblock {\em J. Comput. Sys. Sci.}, 55(1):24--35, 1997.

\bibitem{shaltielumans}
R.~Shaltiel and C.~Umans.
\newblock Pseudorandomness for approximate counting and sampling.
\newblock In {\em Proc. IEEE Conference on Computational Complexity}, 2005.
\newblock To appear.

\bibitem{shamir}
A.~Shamir.
\newblock {IP=PSPACE}.
\newblock {\em J. ACM}, 39(4):869--877, 1992.

\bibitem{shannon}
C.~Shannon.
\newblock The synthesis of two-terminal switching circuits.
\newblock {\em Bell System Technical Journal}, 28(1):59--98, 1949.

\bibitem{shi:gate}
Y.~Shi.
\newblock Both {T}offoli and controlled-{NOT} need little help to do universal
  quantum computation.
\newblock {\em Quantum Information and Computation}, 3(1):84--92, 2002.
\newblock quant-ph/0205115.

\bibitem{sipser:bpp}
M.~Sipser.
\newblock A complexity theoretic approach to randomness.
\newblock In {\em Proc. ACM STOC}, pages 330--335, 1983.

\bibitem{stockmeyer}
L.~J. Stockmeyer.
\newblock The complexity of approximate counting.
\newblock In {\em Proc. ACM STOC}, pages 118--126, 1983.

\bibitem{sm}
L.~J. Stockmeyer and A.~R. Meyer.
\newblock Cosmological lower bound on the circuit complexity of a small problem
  in logic.
\newblock {\em J. ACM}, 49(6):753--784, 2002.

\bibitem{toda}
S.~Toda.
\newblock {PP} is as hard as the polynomial-time hierarchy.
\newblock {\em SIAM J. Comput.}, 20(5):865--877, 1991.

\bibitem{umans:thesis}
C.~Umans.
\newblock {\em Approximability and Completeness in the Polynomial Hierarchy}.
\newblock PhD thesis, UC Berkeley, 2000.

\bibitem{valiant}
L.~G. Valiant.
\newblock The complexity of computing the permanent.
\newblock {\em Theoretical Comput. Sci.}, 8(2):189--201, 1979.

\bibitem{vv}
L.~G. Valiant and V.~V. Vazirani.
\newblock {NP} is as easy as detecting unique solutions.
\newblock {\em Theoretical Comput. Sci.}, 47(3):85--93, 1986.

\bibitem{vinodchandran}
N.~V. Vinodchandran.
\newblock A note on the circuit complexity of {PP}.
\newblock ECCC TR04-056, 2004.

\bibitem{wilson}
C.~B. Wilson.
\newblock Relativized circuit complexity.
\newblock {\em J. Comput. Sys. Sci.}, 31(2):169--181, 1985.

\end{thebibliography}

\section{Appendix: A \textit{Really} Big Crunch\label{CRUNCH}}

By slightly modifying the construction of Theorem \ref{ppthm}, we can resolve
two other open questions of Fortnow.

\begin{theorem}
\label{trifecta}\quad

\begin{enumerate}
\item[(i)] There exists an oracle relative to which $\mathsf{P}^{\mathsf{NP}%
}=\mathsf{PEXP}$, and indeed $\mathsf{P}^{\mathsf{NP}}=\mathsf{P}%
^{\mathsf{NP}^{\mathsf{PEXP}}}$.

\item[(ii)] There exists an oracle relative to which $\mathsf{\oplus
P}=\mathsf{PEXP}$.
\end{enumerate}
\end{theorem}

\begin{proof}
\quad

\begin{enumerate}
\item[(i)] In the oracle construction of Theorem \ref{ppthm} dealing with all
$n$ simultaneously, make the following simple change. \ Whenever a row $R$
gets encoded, record the \textquotedblleft current time\textquotedblright%
\ $t$\ as a prefix to that row. \ In other words, the oracle $A$ will now take
two kinds of queries: those of the form $\left\langle R,i,x\right\rangle $ as
before, and those of the form $\left\langle R,j\right\rangle $ for an integer
$j\geq0$.\ \ Initially $A\left(  R,j\right)  =0$ for all $R,j$. \ At any step
of the iterative procedure, let $t$ be the number of encoding steps that have
already occurred. \ Then call the pair $\left\langle i,x\right\rangle $
\textquotedblleft sensitive\textquotedblright\ to row $R$, if there exists an
oracle $A^{\prime}$\ such that

\begin{itemize}
\item $A^{\prime}$ disagrees with $A$ only in row $R$,

\item $M_{i,x}\left(  A^{\prime}\right)  \neq M_{i,x}\left(  A\right)  $, and

\item as we range over $j$, the $A^{\prime}\left(  R,j\right)  $'s encode the
binary expansion of $t+1$.
\end{itemize}

Clearly the proof of Theorem \ref{ppthm}\ still goes through with this change.
\ For let $\ell=\left\lceil \log_{2}n\right\rceil $. \ Then as before,
whenever there does not exist a row $R$\ of the form $\left(  r_{1}^{\ast
},\ldots,r_{\ell-1}^{\ast},r_{\ell}\right)  $ to which no $\left\langle
i,x\right\rangle $ is sensitive, we can encode a subset of those rows so as to
double $Q\left(  A\right)  $. \ Since $2^{-2^{O\left(  n\right)  }}\leq
Q\left(  A\right)  \leq2^{2^{O\left(  n\right)  }}$\ for all $A$, this process
will halt after at most $2^{O\left(  n\right)  }$\ steps, meaning that
$t$\ will never require more than $O\left(  n\right)  $\ bits to represent.
\ Indeed, this is true even if we are dealing with $\mathsf{PTIME}\left(
2^{n}\right)  $\ machines, rather than $\mathsf{PTIME}\left(  n^{\log
n}\right)  $\ machines.

Now consider a $\mathsf{PTIME}^{A}\left(  2^{n}\right)  $\ machine $M_{i}%
$.\ \ We can simulate $M_{i}$\ in $\mathsf{DTIME}\left(  n^{2}\right)
^{\mathsf{NP}^{A}}$, as follows.\ \ Given an input $x\in\left\{  0,1\right\}
^{n}$, first find the unique row $R=\left(  r_{1},\ldots,r_{\left\lceil
\log_{2}n\right\rceil }\right)  $\ for which $t$\ is maximal---in other words,
the last such row to have been encoded. \ This requires $O\left(  n\right)
$\ adaptive queries to the $\mathsf{NP}$\ oracle, each of size $O\left(
n\right)  $. \ Then output $A\left(  R,i,x\right)  $.

It follows that $\mathsf{DTIME}\left(  n^{2}\right)  ^{\mathsf{NP}%
}=\mathsf{PE}$\ relative to $A$, and (by padding) that $\mathsf{P}%
^{\mathsf{NP}}=\mathsf{PEXP}$. \ Indeed, once the $\mathsf{P}^{\mathsf{NP}}%
$\ machine finds the $r_{\ell}$'s, it can use them to decide an arbitrary
language in $\mathsf{P}^{\mathsf{NP}^{\mathsf{PEXP}}}$, which is why
$\mathsf{P}^{\mathsf{NP}}=\mathsf{P}^{\mathsf{NP}^{\mathsf{PEXP}}}$\ as well.

\item[(ii)] In this case the change to Theorem \ref{ppthm}\ is even simpler.
\ Whenever we encode a row $R=\left(  r_{1},\ldots,r_{\ell}\right)  $, instead
of setting $A_{t}\left(  R,i,x\right)  :=M_{i,x}\left(  A_{t-1}\right)  $ for
all $i,x$, we now set%
\[
A_{t}\left(  R,i,x\right)  :=M_{i,x}\left(  A_{t-1}\right)  \oplus%
{\displaystyle\bigoplus\limits_{R^{\prime}\neq R}}
A_{t}\left(  R^{\prime},i,x\right)  ,
\]
where the sum mod $2$ ranges over all $R^{\prime}=\left(  r_{1}^{\prime
},\ldots,r_{\ell}^{\prime}\right)  $\ other than $R$ itself. \ Then when we
are done, by assumption $A$ will satisfy%
\[
M_{i,x}\left(  A\right)  =\bigoplus_{R=\left(  r_{1},\ldots,r_{\ell}\right)
}A\left(  R,i,x\right)
\]
for all $n\leq2^{\ell}$, $i\in\left\{  1,\ldots,n\right\}  $, and
$x\in\left\{  0,1\right\}  ^{n}$. \ So to simulate a $\mathsf{PE}$\ machine
$M_{i}$\ on input $x$, a $\mathsf{\oplus DTIME}\left(  n\right)  $\ machine
just needs to return the above sum. \ Hence $\mathsf{\oplus DTIME}^{A}\left(
n\right)  =\mathsf{PE}^{A}$, and $\mathsf{\oplus P}^{A}=\mathsf{PEXP}^{A}$\ by padding.
\end{enumerate}
\end{proof}

\end{document}